\documentclass{ieeeaccess}
\usepackage{cite}
\usepackage{amsmath,amssymb,amsfonts}
\usepackage{algorithmic}
\usepackage{graphicx}
\usepackage{textcomp}

\usepackage{bm}

\usepackage{booktabs}                  
\usepackage{xspace}

\newcommand{\ie}{i.e.\@\xspace} 

\newcommand{\commentout}[1]{}

\usepackage{spverbatim}
\usepackage{enumitem}
\usepackage{array}

\usepackage{color}
\usepackage{framed}
\usepackage{listings}
\usepackage{xcolor}

\usepackage{xcolor}    
\usepackage{environ}   

\NewEnviron{myverbbox}{%
  \par\noindent
  \begingroup
    \setlength{\fboxsep}{0pt}%
    \setlength{\fboxrule}{0pt}%
    \colorbox{gray!10}{%
      \begin{minipage}{\linewidth}%
        \vspace{2mm}
        \ttfamily     
        \setlength{\parindent}{0pt}
        \setlength{\leftskip}{2mm}
        \setlength{\rightskip}{2mm}
        \BODY         
        \par\vspace{2mm}
      \end{minipage}%
    }%
  \endgroup
  \par
}

\NewSpotColorSpace{PANTONE}
\AddSpotColor{PANTONE} {PANTONE3015C} {PANTONE\SpotSpace 3015\SpotSpace C} {1 0.3 0 0.2}
\SetPageColorSpace{PANTONE}
\definecolor{accessblue}{cmyk}{1, 0.3, 0, 0.2}
\definecolor{greycolor}{cmyk}{0,0,0,.8}

\makeatletter
\AtBeginDocument{\DeclareMathVersion{bold}
\SetSymbolFont{operators}{bold}{T1}{times}{b}{n}
\SetSymbolFont{NewLetters}{bold}{T1}{times}{b}{it}
\SetMathAlphabet{\mathrm}{bold}{T1}{times}{b}{n}
\SetMathAlphabet{\mathit}{bold}{T1}{times}{b}{it}
\SetMathAlphabet{\mathbf}{bold}{T1}{times}{b}{n}
\SetMathAlphabet{\mathtt}{bold}{OT1}{pcr}{b}{n}
\SetSymbolFont{symbols}{bold}{OMS}{cmsy}{b}{n}
\renewcommand\boldmath{\@nomath\boldmath\mathversion{bold}}}
\makeatother

\def\BibTeX{{\rm B\kern-.05em{\sc i\kern-.025em b}\kern-.08em
    T\kern-.1667em\lower.7ex\hbox{E}\kern-.125emX}}

\begin{document}
\history{Date of publication xxxx 00, 0000, date of current version xxxx 00, 0000.}
\doi{10.1109/ACCESS.2024.0429000}

\title{MagicCraft: Natural Language-Driven Generation of Dynamic and Interactive 3D Objects for Commercial Metaverse Platforms}
\author{\uppercase{Ryutaro Kurai}\authorrefmark{1,2}, \IEEEmembership{Student Member, IEEE},
\uppercase{Takefumi Hiraki}\authorrefmark{3,4}, \IEEEmembership{Member, IEEE},\\
\uppercase{Yuichi Hiroi}\authorrefmark{3}, \IEEEmembership{Member, IEEE},
\uppercase{Yutaro Hirao}\authorrefmark{2},
\uppercase{Monica Perusqu\'{i}a-Hern\'{a}ndez}\authorrefmark{2},
\uppercase{Hideaki Uchiyama}\authorrefmark{2},
and \uppercase{Kiyoshi Kiyokawa}\authorrefmark{2}, \IEEEmembership{Member, IEEE}
}

\address[1]{Cluster, Inc., 8-9-5 Nishogotanda, Shingawa, Tokyo 141-0031 Japan (e-mail: r.kurai@cluster.mu)}
\address[2]{Graduate School of Science
and Technology, Nara Institute of Science and Technology, 8916-5 Takayama, Ikoma, Nara 630-0192 Japan (e-mail: \{yutaro.hirao, m.perusquia, hideaki.uchiyama, kiyo\}@is.naist.jp)}
\address[3]{Cluster Metaverse Lab, 8-9-5 Nishogotanda, Shingawa, Tokyo 141-0031 Japan (e-mail: \{t.hiraki, y.hiroi\}@cluster.mu)}
\address[4]{Institute of Library, Information and Media Science, University of Tsukuba, 1-2 Kasuga, Tsukuba, Ibaraki 305-8550 Japan}
\tfootnote{This work was partially supported by JST ASPIRE Grant Number JPMJAP2327. (This is a preprint version)}

\markboth
{Kurai \headeretal: MagicCraft: Natural Language-Driven Generation of Dynamic and Interactive 3D Objects for Commercial Metaverse Platforms}
{Kurai \headeretal: MagicCraft: Natural Language-Driven Generation of Dynamic and Interactive 3D Objects for Commercial Metaverse Platforms}

\corresp{Corresponding author: Ryutaro Kurai (e-mail: r.kurai@cluster.mu).}

\begin{abstract}
Metaverse platforms are rapidly evolving to provide immersive spaces for user interaction and content creation. However, the generation of dynamic and interactive 3D objects remains challenging due to the need for advanced 3D modeling and programming skills. To address this challenge, we present MagicCraft, a system that generates functional 3D objects from natural language prompts for metaverse platforms. MagicCraft uses generative AI models to manage the entire content creation pipeline: converting user text descriptions into images, transforming images into 3D models, predicting object behavior, and assigning necessary attributes and scripts. It also provides an interactive interface for users to refine generated objects by adjusting features such as orientation, scale, seating positions, and grip points.

Implemented on Cluster, a commercial metaverse platform, MagicCraft was evaluated by 7 expert CG designers and 51 general users. Results show that MagicCraft significantly reduces the time and skill required to create 3D objects. Users with no prior experience in 3D modeling or programming successfully created complex, interactive objects and deployed them in the metaverse. Expert feedback highlighted the system's potential to improve content creation workflows and support rapid prototyping. By integrating AI-generated content into metaverse platforms, MagicCraft makes 3D content creation more accessible.
\end{abstract}


\titlepgskip=-21pt

\maketitle

\section{Introduction}
\begin{figure*}
  \includegraphics[width=\textwidth]{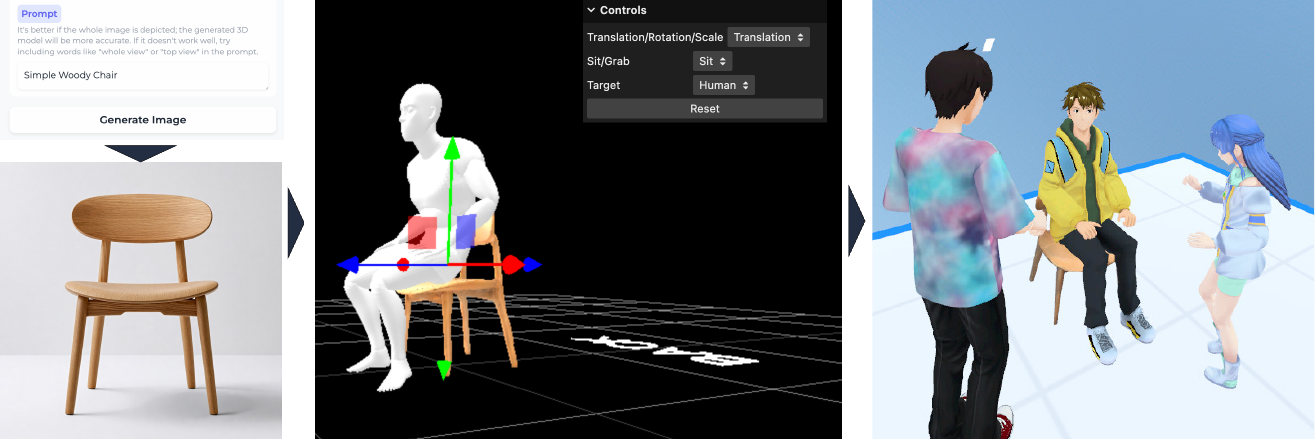}
  \caption{When the user describes the object to be created in natural language, an image of the object is generated. A 3D object is then created from the generated image. The user sets the behavior, such as the position where the object will sit. The object with the set behavior is immediately uploaded to the metaverse space and can be used by multiple people.}
  \label{fig:teaser}
\end{figure*}

\PARstart{T}{he} emergence of metaverse platforms, driven by rapid advancements in virtual reality (VR) technology, now provides immersive spaces where users interact, collaborate, and express creativity through avatars and virtual content. Platforms such as VRChat~\cite{vrchat}, Roblox~\cite{roblox}, and Cluster~\cite{clsuter} have attracted millions of users around the world through their rich 3D virtual environments.

User-generated content (UGC), specifically the creation of interactive 3D objects, is at the heart of the metaverse experience. However, the development of such objects requires significant expertise in both 3D modeling and programming, skills that present significant barriers to non-experts. This limitation limits creative participation and diminishes the diversity of content essential to the growth of the metaverse. Our research addresses this accessibility gap through an integrated interface that leverages generative AI capabilities to enable users without technical expertise to create functional objects for metaverse environments.

Recent advances in large language models (LLMs) and generative AI offer promising approaches to content creation. Models such as Stable Diffusion~\cite{rombach2022stable} generate 2D images from text descriptions, while DreamFusion~\cite{poole2023dreamfusion} generates 3D objects from textual or visual input. In addition, multimodal LLMs such as OpenAI GPT-4~\cite{OpenAI_GPT4_2023} can recognize visual media and generate code, including scripts that define object behavior~\cite{Giunchi2024-gt, kurai-2025-magicitem}.

However, the integration of these generative technologies into metaverse platforms presents several challenges. First, ensuring compatibility between generated 3D objects and platform specifications requires navigating differences in file formats, rendering pipelines, and interaction mechanisms. Second, translating natural language into functional 3D models with precise behaviors requires coordination among multiple AI models, each with different limitations in image fidelity and 3D reconstruction accuracy. Third, defining interactions between generated objects and users involves inherent ambiguities because there is no universal mapping between objects and interaction patterns.

We propose MagicCraft, a system that enables users to generate functional 3D objects for metaverse platforms from natural language prompts. MagicCraft orchestrates a complete content creation pipeline: generating images from textual descriptions, converting these into 3D models, predicting object behaviors, and assigning appropriate attributes and scripts. The system also provides an interactive interface to refine the generated objects, allowing adjustments to the orientation, scale, seating positions, and grip points.
We implemented MagicCraft on Cluster, a commercial metaverse platform with more than 35 million users. By integrating with Cluster's API and scripting language, users can upload and manipulate generated objects directly within the platform.

Building on our previous work~\cite{kurai-2025-IEEEVR-poster}, we conducted a comprehensive evaluation to assess the usability of the system and quantify its time efficiency compared to traditional object creation methods. Our evaluation combines a comparative analysis with seven expert CG designers and an online experiment with 51 general users. The results show that MagicCraft significantly reduces object creation time, while lowering technical barriers to entry. Even participants with no 3D modeling or programming experience successfully created complex, interactive objects and deployed them in metaverse environments. Expert evaluations confirmed the usefulness of the system for rapid prototyping and improving content creation workflows.

MagicCraft represents a significant step forward in democratizing 3D content creation by addressing key challenges in integrating AI-generated content into metaverse platforms. As generative AI continues to evolve, systems like MagicCraft can foster more inclusive and dynamic metaverse ecosystems. Key contributions of this research include:

\begin{itemize}[leftmargin=*]
\item Developing MagicCraft, A natural language-based 3D object generation tool that integrates LLMs and generative models on a large-scale commercial metaverse platform.
\item Enabling the production of dynamic objects with accurate scaling and interaction features through the integration of image-to-3D conversion and autonomous behavior scripting.
\item Demonstrating the system's effectiveness in reducing creation time and lowering skill barriers, validated by expert and general user feedback in a large-scale user study.
\item Providing insights into current challenges and outlining pathways for further research in AI-assisted 3D content creation for the metaverse.
\end{itemize}

\section{Related Work}
\subsection{Metaverse Platforms and 3D Content Creation}
Metaverse platforms such as VRChat~\cite{vrchat}, Roblox~\cite{roblox}, Neos~\cite{neos}, Resonite~\cite{resonite}, and Cluster~\cite{clsuter} have emerged as virtual environments where users can interact, collaborate, and create content in immersive 3D spaces. 
A central feature of these platforms is UGC, which empowers users to design and share virtual worlds, objects, and experiences~\cite{Ondrejka2004}. Despite the creative opportunities, creating dynamic and interactive 3D objects is a significant challenge for many users due to the complexity of 3D modeling software and the need for programming skills for interactivity.
Creating 3D models typically requires mastery of advanced tools such as Blender~\cite{blender}, which have steep learning curves and require significant time investment. In addition, implementing interactive behaviors requires an understanding of programming languages and scripting within game engines or platform-specific environments, such as Unity~\cite{unity} and Unreal Engine~\cite{unreal-engine}. These barriers not only limit participation to users with specialized skills, but also limit the diversity of content within metaverse platforms~\cite{cho2015physical}.

Research in virtual environments has examined various aspects of user interaction and content creation. Studies have examined the impact of avatar representation on user experience~\cite{Latoschik2017}, the role of nonverbal communication in virtual social interactions~\cite{Pan2017}, and the ways in which users engage with spatial relationships and objects~\cite{Hindmarsh2000}. While this work underscores the importance of accessible tools to enhance user engagement, the challenge of democratizing the creation of interactive 3D content remains~\cite{bowman2008ui}. Addressing this issue is critical to fostering inclusive participation and enriching the metaverse ecosystem.

\subsection{Generative AI for Text-to-3D Object Creation and Behavior Modeling}
Advances in generative AI technology have led to novel methods to create 3D models from text descriptions, with the goal of lowering the barriers to the creation of 3D content~\cite{poole2023dreamfusion, tang2023dreamgaussian, bensadoun2024meta3dgen, SF3D-Boss2024-uz, lin2023magic3d}.
Despite these innovations, integrating AI-generated models into interactive virtual environments presents several challenges. Generative models often produce 3D assets that lack the necessary structure or optimization for real-time rendering and interaction~\cite{Xie2022}. In addition, the generated objects may not meet the compatibility requirements of specific metaverse platforms, such as polygon count limitations, texture formats, animation rigging~\cite{huynhthe2023ai}.

Text-to-3D models typically focus on static geometry and do not inherently provide interactive behaviors or scripts that define how objects should respond in the environment.
To address behavior modeling, AI-based systems have been developed to automate the generation of object interactions and animations. 
For example, MotionDiffuse~\cite{zhang2022motiondiffuse} synthesizes avatar motion sequences based on textual input. Although these systems enhance the realism and interactivity of virtual scenes, they often require detailed descriptions and may not scale well to different object categories or complex behaviors.

LLMs have been used to generate code for behavior scripting. Platforms such as LLMR~\cite{De_La_Torre2024-be} and DreamCodeVR~\cite{Giunchi2024-gt} allow users to enter natural language commands that are then converted into code scripts that control object behaviors within game engines. While these approaches facilitate the creation of interactive content, they often rely on users to provide precise instructions and may produce code that requires debugging or optimization. In addition, ensuring that the generated code matches the intent of the user and works securely within the platform remains a challenge~\cite{Pearce2022}.

\subsection{User Interfaces and LLMs for Accessible 3D Content Creation}
Developing intuitive user interfaces is essential to make AI-generated content accessible to users without technical experience. Interfaces that allow users to interactively refine generated models bridge the gap between automated generation and personalized content~\cite{koyama2022bo}.

In the context of behavior scripting, user interfaces that abstract the complexity of programming can enable users to define object interactions more intuitively. Visual scripting tools, such as Unreal Engine's Blueprint or Unity's Visual Scripting, provide node-based interfaces for creating scripts without writing code. However, these tools still require an understanding of the logic flow and programming concepts that may not be accessible to all users.

LLMs trained in code repositories have shown a significant ability to generate code from natural language prompts~\cite{Chen2021}. Tools such as GitHub Copilot~\cite{github-copilot} assist developers in suggesting code snippets and functions based on comments or code snippets. Although these tools increase developer productivity, their effectiveness is limited for users without programming background~\cite{Vaithilingam2022}. Integrating LLMs into user-friendly interfaces that guide users through content creation processes can make advanced functionality more accessible~\cite{ahmad-etal-2021-unified}.

The challenge of ensuring that AI-generated content meets the quality, security, and compatibility standards required for use in metaverse platforms remains. In addition, differences in platform-specific requirements such as scripting languages, interaction models, and asset pipelines require adaptable solutions that can tailor output to the target environment. Overcoming these challenges is critical for systems like MagicCraft to effectively enable users to create and deploy interactive 3D content with confidence.

\section{MagicCraft}
\begin{figure*}[t]
    \centering
    \includegraphics[width=1\linewidth]{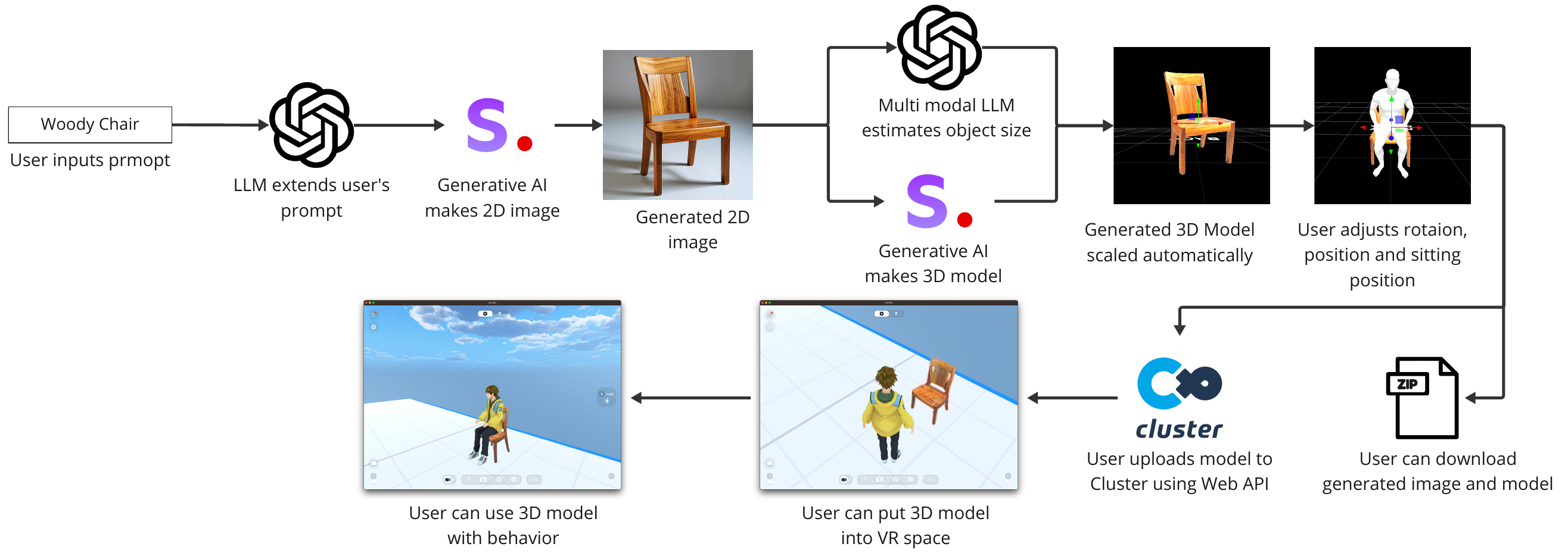}
    \vspace*{-3mm}
    \caption{The initial user input is augmented by a large language model (LLM), and this augmented input is used as input for an AI to generate an image. The resulting image is used as input for an AI to generate a 3D model, and also helps to estimate the real-world size of the object. The 3D model is scaled based on the estimated size, and users can adjust the object's behavior within the system. The finished object can be immediately uploaded to metaverse services, where users can place or interact with it in virtual space. Users can also download the object as a local file.}
    \label{fig:system_overvies}
    \vspace{-3mm}
\end{figure*}

The primary goal of MagicCraft is to enable users to generate 3D objects and use them in metaverse spaces by simply entering natural language descriptions, \ie, prompts.
MagicCraft lies in its ability to perform the entire process, from natural language input to 3D object generation, customization, and integration into metaverse spaces -- consistently on a single platform. 

First, we describe the overall workflow that the user performs through MagicCraft. Then the detailed design of each component is described.

Note that although we used specific generative AI models and the metaverse platform in the implementation, the system design described in this section is applicable to any environment. 

\subsection{System Overview}
Figure~\ref{fig:system_overvies} shows an overview of MagicCraft. First, the user imagines the object he wants to create and enters it as a prompt. The system then generates a 2D image based on this prompt. Users can review this image and iterate through the image generation process until the desired object is accurately represented. Once satisfied with the 2D representation, the user can proceed to generate a 3D object from this image. Following the system's guidance, the user can adjust the orientation, scale, and behavior of the generated object. Finally, the finished 3D object can be uploaded directly to a metaverse platform, allowing users to immediately observe and interact with the object in the metaverse environment. This streamlined process enables rapid prototyping and deployment of user-generated content in virtual spaces.

\subsection{Component Design}\label{sec:component-design}
\subsubsection{Image Generation} 
Although users tend to input short instructions of 2--3 words, image generation is known to use more detailed instructions, resulting in more complex and physically accurate images of the shape.
Therefore, prior to image generation, user input prompts are first sent to the LLM to provide improved prompts tuned for image generation. Specifically, the following instructions are given to the LLM:

\vspace{1mm}
\begin{myverbbox}
Generate an English prompt for image creation. Please make it a short statement. The prompt should contain the following conditions: \\
\# Conditions \\
* Describe the overall image of the subject. \\
* Capture the subject from the front. \\
* Do not include any background. \\
* Draw it as if it were a photograph.
\end{myverbbox}

Using this enhanced prompt as input to the image generation model, concise user input is used to generate detailed and accurate images suitable for output as 3D models in the metaverse platform.

Note that a generative AI model has recently been proposed that outputs a 3D model directly from the input prompts. However, once the output is in the form of an image, the user can refine the input prompt before generating the 3D model, reducing the time required for trial and error. It would also allow users to directly upload images and convert them to 3D.

\subsubsection{Scale-Aware 3D Object Generation from Images}
We convert the generated image into a 3D object using image-to-3D generative AI models. A major challenge in this process is that these models typically standardize the output objects to approximately 1 m$^3$ volumes, ignoring the actual proportions of the represented object. To overcome this limitation, we have implemented an automatic scaling mechanism that preserves real-world proportions.

Our scaling pipeline consists of three main steps. First, we used a multimodal LLM to analyze the generated image and estimate the dimensions of the object in real world contexts. 

The model is given the following prompt:

\vspace{1mm}
\begin{myverbbox}
You are an excellent surveyor. Please estimate the actual length of the longest part of the object appearing in the image. \\
For example, if the object shown in the image is a dog, estimate its total length - from the tip of the tail to the tip of the nose.
\end{myverbbox}

Second, we extract the dimensional boundaries from the generated GLTF file \cite{GLTF-The-Khronos(r)-3D-Formats-Working-GroupUnknown-uc} by calculating the maximum and minimum coordinates along each axis. We define the computational length of the object as the largest span across each dimension.

Finally, we compute the scaling factor by determining the ratio of the real-world length estimated by LLM to the computational length of the object. This factor is applied to the 3D model, resulting in objects with proportions that match real-world expectations.

The implementation leverages multimodal LLMs specifically because they incorporate general knowledge about object dimensions, a critical requirement for handling the diverse range of objects that users may create.
This automatic scaling approach allows MagicCraft to generate 3D objects with dimensionally appropriate representations, supporting placement and interaction in metaverse environments.

\subsubsection{System-Assisted Object Adjustment}
MagicCraft provides functionality to enable behavioral settings necessary for avatar interaction, thereby enhancing the functionality of objects in metaverse spaces. In particular, it is essential to configure settings that allow avatars to perform basic actions such as sitting on or grasping objects.
On the other hand, there are also user preferences in setting up such behaviors. For example, different users may have different preferences for sitting astride or sideways against the horse's back.

\begin{figure}[t]
    \centering
    \includegraphics[width=1\linewidth]{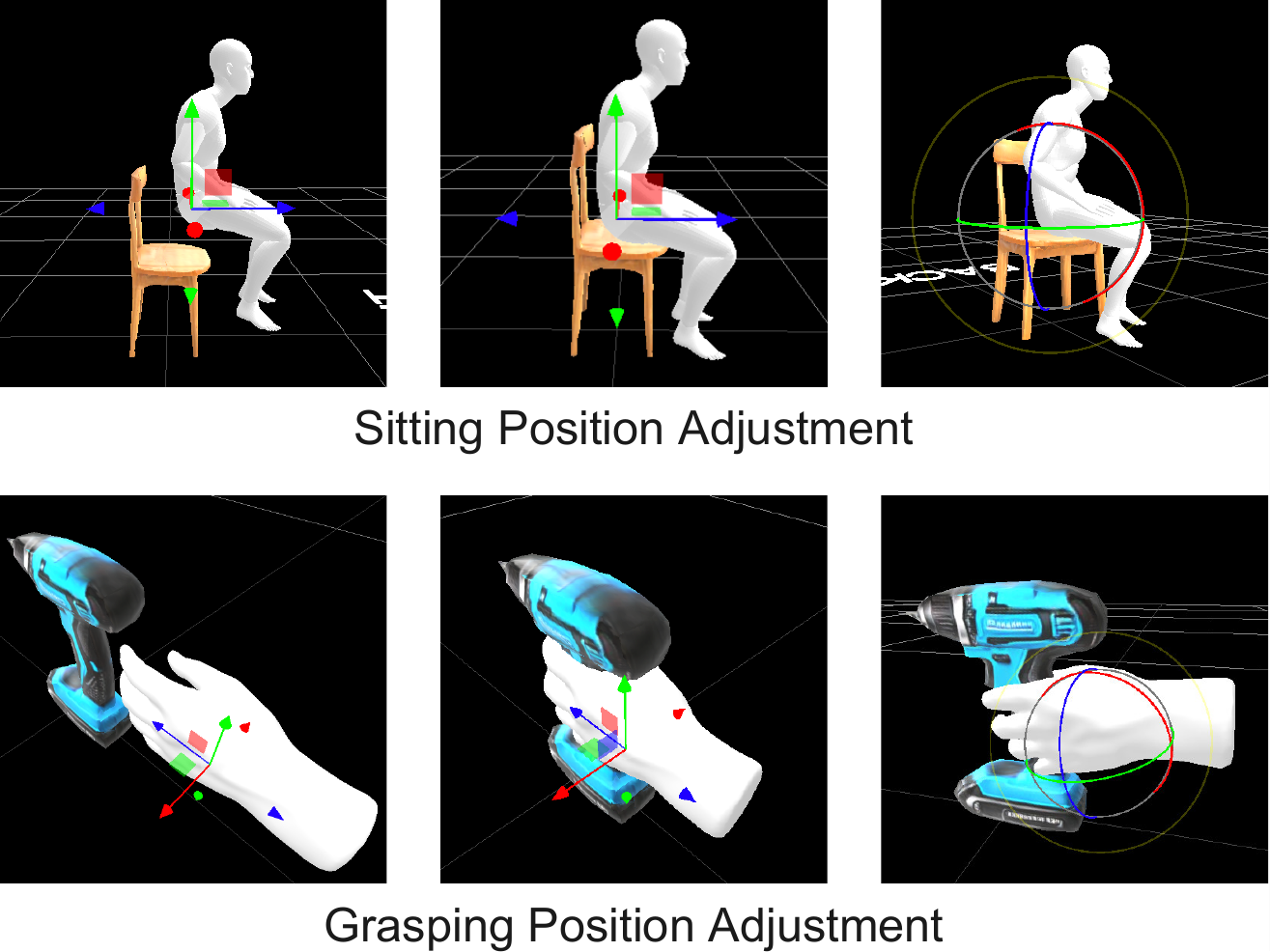}
    \caption{Position Adjustment User Interfaces}
    \label{fig:position_adjustment}
\end{figure}

To address this problem, our system presents a reasonable initial position while providing a user interface that allows the user to adjust the sitting and grasping positions (Fig.~\ref{fig:position_adjustment}). Specifically, by displaying a 170 cm height mannequin object on the 3D interface and adjusting the relative positions of this mannequin and the generated object, the user determines the avatar's interaction with the objects in the metaverse space.
These behavioral settings allow the MagicCraft system to use user-created 3D objects more naturally and purposefully in metaverse space. 

\subsubsection{Autonomous Scripting for Virtual Objects}
This study presents an innovative approach to impart autonomous behavior to objects in metaverse environments. Traditionally, programming complex object motions and rotations has required extensive programming expertise. For example, implementing the flight dynamics of an aircraft requires precise definition of multiple parameters such as altitude, speed, and orientation.

Our MagicCraft system overcomes this challenge by automatically generating appropriate behavior scripts from initial object images. 
While there are several approaches to generating scripts with the LLM, we used a method where the script definitions from the metaverse platform are added directly to the input prompts and entered into the LLM. The generated scripts are retained in the system and subsequently integrated with their corresponding objects during the assembly phase.

\subsubsection{Assembly and Upload Process}
Our system generates several types of data from user input. These include 2D images, 3D models (consisting of meshes and textures), 3 vectors defining object position, orientation, and scale, 2 vectors indicating sitting or grasping positions and orientations, and behavior-defining scripts. We refer to the process of consolidating these heterogeneous data types into a single file called as assemply. This assembly is formatted according to the format of each metaverse platform and uploaded directly to the metaverse platform via web API. The user can then view this object, including its behavior, on the platform.

\begin{figure}[t]
    \centering
    \includegraphics[width=1\linewidth]{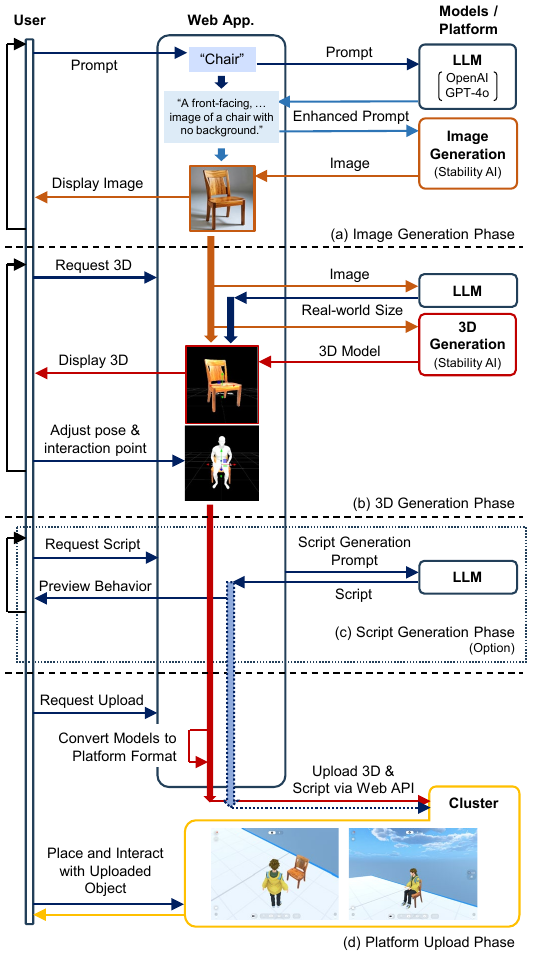}
    \caption{MagicCraft system architecture illustrating the end-to-end workflow across four sequential phases: (a) image generation phase, (b) 3D generation phase, (c) optional script generation phase for behavioral programming, and (d) platform upload phase for cluster integration. Data flows are shown between user interactions (left), web application processing (center), and specialized AI services and platform integration (right).}
    \label{fig:system-diagram}
\end{figure}

\section{Implementation}\label{sec:implementation}
This section describes the implementation of MagicCraft on the commercial metaverse platform Cluster. As shown in Figure~\ref{fig:system-diagram}, the system architecture performs the conceptual phases described in Sec.~\ref{sec:component-design} through a web-based interface that orchestrates multiple specialized services.
The implementation follows four phases: (a) image generation phase, where user instructions are enhanced and rendered as 2D images; (b) 3D generation phase, which includes automatic scale estimation and interactive adjustment capabilities; (c) optional script generation phase for behavioral programming using LLM-generated script; and (d) platform upload phase for direct integration with Cluster. This architecture allows users to create complex 3D objects without directly managing the underlying API calls or technical integration between components.

\begin{figure*}
    \centering
    \includegraphics[width=1\linewidth]{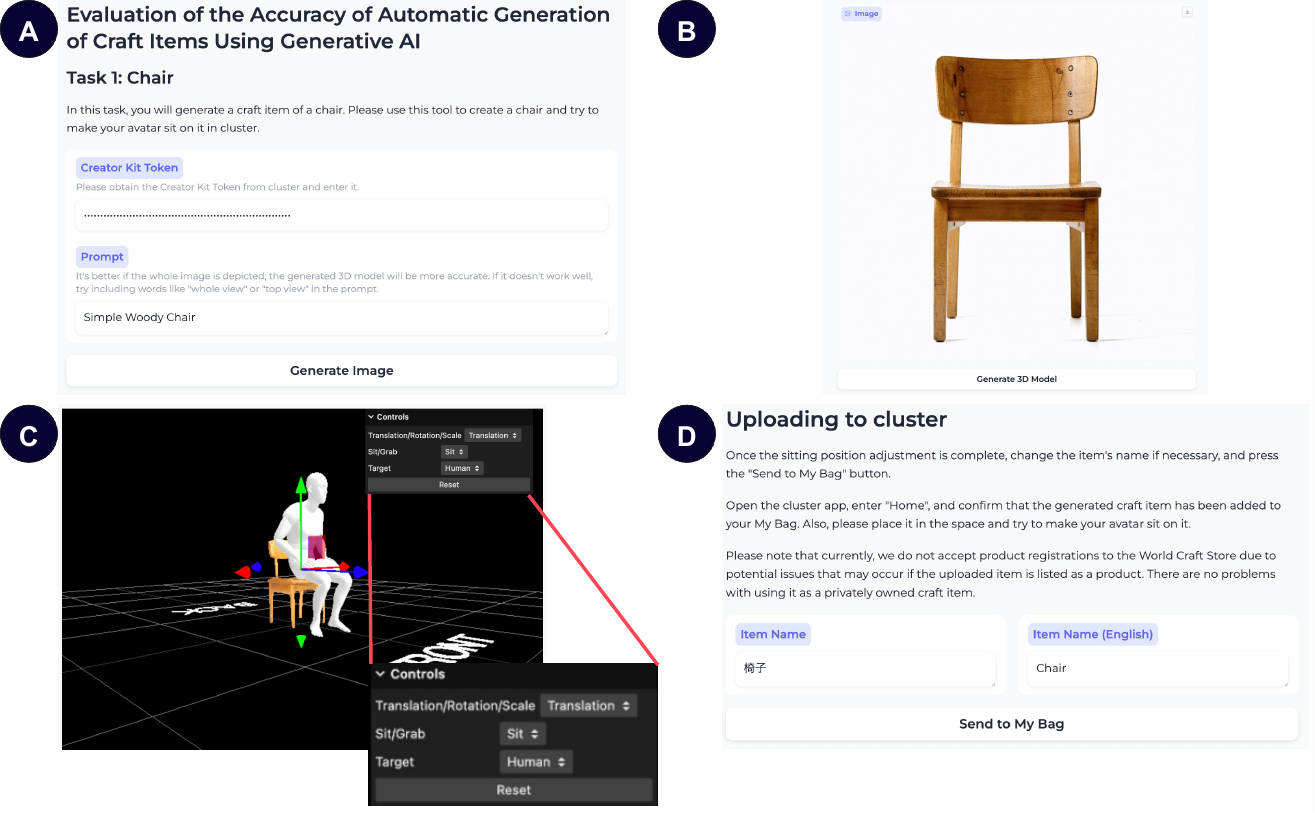}
    \caption{The interface of MagicCraft web application. (A) Initial prompt input screen with API token entry field and text description input for object generation. (B) Generated image preview with options to accept or regenerate. (C) 3D model adjustment interface with controllable parameters for position, rotation, scale, and interactive points (sitting/grabbing). The mannequin provides visual reference for avatar interaction. (D) Upload interface with naming fields and confirmation options for deploying the 3D object to the Cluster metaverse platform.}
    \label{fig:web_interface}
\end{figure*}

\paragraph{Metaverse Platform and Script Generation} We implemented our system on Cluster, a commercial metaverse platform with more than 35 million cumulative users, which facilitated efficient recruitment of participants for our evaluation studies. 

Cluster provides native support for object behavior programming through ClusterScript, a TypeScript-based language that enables runtime execution of interactive behaviors within the client environment. 

For script generation, we have implemented prompts in previous research~\cite{kurai-2025-magicitem} and produced valid ClusterScript code on OpenAI GPT-4o~\cite{GPT-4o-mini} that can be integrated directly into generated objects.

\paragraph{Web Application} 
Figure~\ref{fig:web_interface} shows the appearance of our system. 
To implement our system as a web application, we used Gradio~\cite{GRADIO-Abid2019-sr}, a Python-based web application framework. 

First, users enter their Cluster API key and object description as a natural language prompt (Fig.~\ref{fig:web_interface}, A). After clicking "Generate Image", the system processes this input through the AI pipeline. The interface then displays the resulting image preview, allowing the user to proceed with 3D model generation or iterate on the image by adjusting the prompt or regenerating (Fig.~\ref{fig:web_interface}, B).
The interface then provides an advanced preview of the generated 3D model along with controls for behavioral adjustments (Fig.~\ref{fig:web_interface}, C). Users can modify position, rotation, and scale parameters using drop-down menus, and configure interaction points for sitting or grasping. The interface displays a reference mannequin (170~cm tall) to visualize avatar interactions, with target selection allowing users to adjust either the object itself or the interaction points. Finally, the interface allows for direct upload to Cluster, where users name their creations before deploying them as personal objects (Fig.~\ref{fig:web_interface}, D).

For system stability and scalability, we containerized the application using Docker~\cite{Docker-Merkel2014-iw} and deployed it on Google Cloud Run~\cite{Cloud-Run}, which allows automatic resource allocation based on traffic demand.

\paragraph{Generative Models}  Our system uses several generative AI models. Basically, the models are selected by balancing the speed and quality of generation. However, since the image generation from the prompt affects the 3D model later, we used a model with high generation quality even if the time is slow.

In this implementation, OpenAI GPT-4o-mini~\cite{GPT-4o-mini} was used for the multimodal input LLM for scale estimation from images, OpenAI GPT-4o~\cite{UGPT-4o}  was used for script generation, Stable Image Ultra~\cite{Stability-AI-Models} was used for image generation from prompts, and Stable Fast 3D~\cite{SF3D-Boss2024-uz} was used for image-to-3D generation.

\paragraph{Parallel Access Capacity} The system's capacity for parallel access is primarily constrained by the usage limits of the Stability.ai and OpenAI web APIs. Specifically, Stability.ai imposes a rate limit of 150 requests per 10 seconds, while OpenAI's most restrictive API allows 10,000 requests per minute. Log analysis shows a peak processing rate of 2 requests per second, with consistent response times across all cases.

\paragraph{Assembly and Upload} Since our experimental target platform is Cluster, we chose the GLB file format as the output of the assembly process. GLB, a binary form of the GLTF specification, allows the encapsulation of meshes, textures, and additional extension data into a unified binary file. 

Object position, rotation, and scale are encoded according to the GLTF standards, while extended information such as sitting or grasping positions are included as GLTF extensions. These assembled objects are then uploaded to Cluster via a web API.

User authentication for cluster uploads is managed by CCK tokens, which are required for API access. As a result, our web application prompts users to enter their CCK token at the initial stage of use, effectively limiting system access to those with valid CCK tokens.

\paragraph{Content and Log Storage for Analysis} We implemented an automated content storage mechanism using Google Cloud Storage for images, 3D models, and other generated assets. 

\section{Web-Based Public Experiment}
To evaluate the proposed system, we conducted a public experiment by deploying the system online and recruiting participants via the Internet. 
This study was approved by Cluster Inc. Research Ethics Committee.

Participants accessed the website made by Gradio for this experiment. The website contained an explanatory document that provided an overview of the experiment, the research purpose, methods, privacy and data handling policies, rewards, and contact information for the person in charge. Users could proceed to the subsequent experimental steps by reading all the information and checking a box to indicate their consent to participate.
All user studies were conducted on participants' personal Windows or Mac PCs.

Before going to the actual tasks, to familiarize users with the system, we instructed the participants to enter ``alarm clock'' as a prompt. The participants were then asked to generate a 3D model from the resulting image and upload it to Cluster. Users then placed the alarm clock in the Cluster VR space and observed its appearance. This process allowed users to experience the complete workflow of the tool suite.

To familiarize participants with the system prior to the actual tasks, we asked participants to complete a practice task in which they were asked to explain a series of steps related to using this system. First, participants were instructed to type ``alarm clock'' as a prompt. Participants were then asked to generate a 3D model from the resulting image and upload it to Cluster. The user then placed the alarm clock in Cluster's VR space and observed its appearance. This process allowed the user to experience the full workflow of the tool suite.

Upon completion of the 3D model creation for each task, users evaluated their models within the Cluster platform, rating their satisfaction using a 5-point Likert scale and free-text comments.
After completing all tasks, participants responded to a post-experiment questionnaire administered via Google Forms. This questionnaire included demographic information, an overall usability rating using the System Usability Scale (SUS)~\cite{Brooke1996-kw}, and a perceived workload rating using the NASA Task Load Index (NASA-TLX)~\cite{Nasa1986-vw}. While the original NASA-TLX uses a 100-point scale, we used a 10-point scale (1-10) due to the limitations of Google Forms. The results were then converted to the 100-point scale using the formula $(n-1) \times 100/9$, where $n$ represents the participant's rating.
In addition, participants were asked to provide open-ended responses describing their overall impression of the system used during the experiment.

As compensation for completing all user study tasks and the post-experiment questionnaire, participants were awarded 5,000 Cluster points. These points represent a virtual currency exclusive to the Cluster platform, equivalent to the purchase of one user-generated avatar or five avatar accessories.

\subsection{Tasks}
We asked participants to perform four types of tasks. The items generated by each task interact in some way with avatars and environments in the metaverse space. 

\subsubsection{Task 1: Create a Sittable Item}
In this task, participants were asked to create a chair. No example instructions were provided to encourage free input from the user. Participants were instructed to generate images from their prompts and create a 3D model once they were satisfied with the generated image. After completing the 3D model, users were asked to adjust its orientation, scale, and seating position. After making these adjustments, participants uploaded the model to Cluster and verified it in the VR space. 

\subsubsection{Task 2: Create a Grabbable Item}
This task involved the creation of an electric drill as a grabbable item. The reason for using an electric drill is that the textures of daily-use grabbable items are more complex than those of a hammer or pencil, and cannot be made easily.

As in Task 1, prompt input was unrestricted. The process from image generation to 3D model creation mirrored that of Task 1, but upon completion of the model, users were asked to adjust the position at which the avatar would hold the drill. Participants then uploaded the model to Cluster, verified it in VR space.

\subsubsection{Task 3: Create an Item with Auto-generated Behavior}
Participants were asked to design a sagiall airplane. The procedure from prompt input to 3D model creation remained the same as in the previous tasks. However, upon completion of the model, participants were required to adjust the orientation and scale, and additionally use an LLM to generate a script representing the movement of the airplane. After these steps, participants uploaded the model to Cluster, reviewed it in the VR space.

\subsubsection{Task 4: Free Creation}
In this task, users were given the freedom to create a craft item of their choice. The process from prompt input to 3D model creation remained the same as in the previous tasks. After completing the model, participants were instructed to adjust orientation and scale, and to use the LLM to generate object motion if necessary. They then uploaded the model to Cluster, reviewed it in VR space.

\subsection{Participants Demography}
 A total of 51 people participated in the experiment. While we received 54 responses to the post-experiment questionnaire, we excluded three responses from our analysis for the following reasons: one response lacked the user identification token, one was a duplicate submission, and one participant completed only a portion of the experimental tasks. As a result, we determined the final number of valid participants to be 51.
The demography of the participants is as follows:
\paragraph{Gender} 52.9~\% identified as male (n = 27), 23.5~\% chose not to disclose their gender (n = 12), and 23.5~\% identified as female (n = 12).
\paragraph{Age} The mean age of the participants was 37.67~$\pm$~9.96 years old, with ages ranging from 18 to 56.
\paragraph{Cluster Usage Frequency} The majority of participants (74.5~\%) reported using Cluster daily (n = 38), followed by 19.6~\% who used it approximately 3 times a week (n = 10). A small percentage reported using it once a week (3.9~\%, n = 2) or once a month (1.9~\%, n = 1).
\paragraph{Crafting Frequency} When asked about how frequently they create craft items, 33.3~\% of the participants reported that they do not craft at all (n = 17), while 31.4~\% craft approximately once every six months (n = 16). Other responses included crafting monthly (19.6~\%, n = 10), more than once a week (7.8~\%, n = 4), and once a week (7.8~\%, n = 4).
\paragraph{Cluster Script Writing Frequency} Regarding the frequency of writing Cluster scripts, 51.0~\% of participants indicated that they do not write scripts at all (n = 26), 21.6~\% write scripts approximately once every six months (n = 11), 15.7~\% write scripts monthly (n = 8), and a small proportion write scripts more than once a week (5.9~\%, n = 3) or once a week (5.9~\%, n = 3).
\paragraph{Experience with 3D Modeling} The average years of experience with 3D modeling reported was 1.87 years (SD = 3.63). The majority of participants had little to no experience, with many indicating 0 years of experience.
\paragraph{Use of Image Generation AI} When asked about their use of image generation AI tools like Stable Diffusion, 39.2~\% of participants reported that they do not use such tools (n = 20). Among those who do, the most common usage frequency was once a month (19.6~\%, n = 10), followed by once a week (15.7~\%, n = 8), and every six months (11.8~\%, n = 6). A smaller group reported daily usage (9.8~\%, n = 5) or approximately three times a week (3.9~\%, n = 2).

\subsection{Quantitative Results}
\begin{figure*}
    \centering
    \includegraphics[width=1\linewidth]{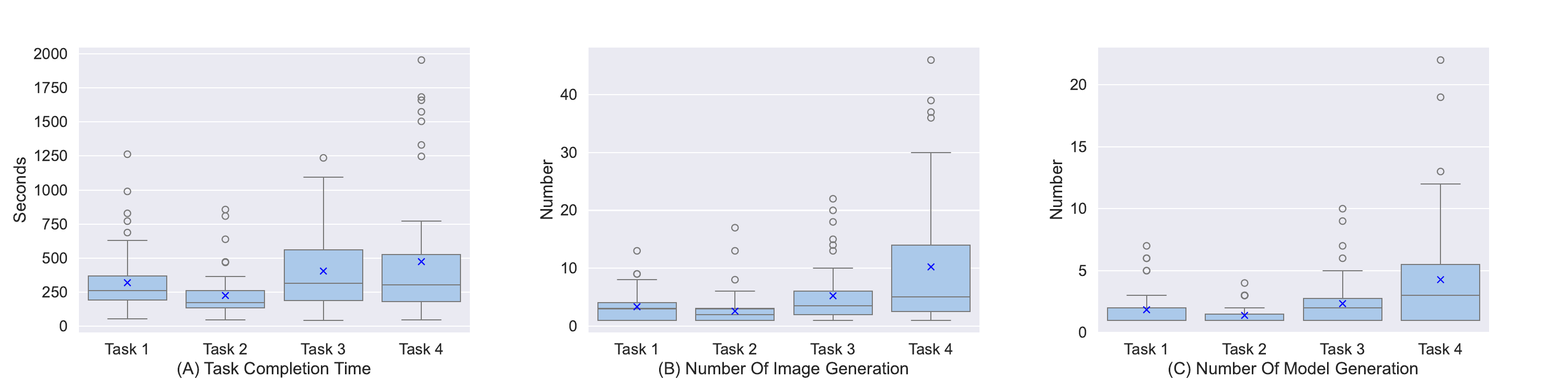}
    \caption{Performance metrics across tasks, including (A) task completion time (in seconds), (B) number of image generations, and (C) number of 3D model generations.}
    \label{fig:generation_analysis}
\end{figure*}

Table~\ref{tab:result_all} shows the results of the online experiment, and Figure~\ref{fig:generation_analysis} shows the time it took the user to complete each task, the number of images generated from the prompts, and the number of 3D models generated from the images.

\begin{table*}[t]
\centering
\caption{Results from online experiments and expert interviews on 3D object creation in the metaverse platform, including task success rates, user satisfaction scores on a 5-point Likert scale, stratified by user experience levels, performance metrics, and estimated time to create target objects based on expert interviews.}
\small
\setlength{\tabcolsep}{4pt}
\begin{tabular}{p{5cm} >{\centering\arraybackslash}p{2.5cm} >{\centering\arraybackslash}p{2.5cm} >{\centering\arraybackslash}p{2.5cm} >{\centering\arraybackslash}p{2.5cm}}
\toprule
& \multicolumn{4}{c}{Task} \\
\cmidrule(lr){2-5}
& 1 & 2 & 3 & 4 \\
\midrule
\multicolumn{5}{l}{\textbf{Task Success Rate}} \\[1ex]
\# of participants & 51 & 51 & 51 & 51 \\[0.5ex]
\# of task success & 51 & 51 & 50 & 50 \\[0.5ex]
\% of Success Rate & 100.0 & 100.0 & 98.0 & 98.0 \\[1ex]
\midrule
\multicolumn{5}{l}{\textbf{Performance Metrices}} \\[1ex]
Task completion time (sec) & 319.8 $\pm$ 238.9 & 226.3 $\pm$ 168.2 & 405.4 $\pm$ 281.7 & 474.7 $\pm$ 480.2\\[0.5ex]
\# of Image Generation & 3.32 $\pm$ 2.67 & 2.57 $\pm$ 2.93 & 5.22 $\pm$ 5.07 & 10.2 $\pm$ 11.2 \\[0.5ex]
\# of 3D Generation & 1.84 $\pm$ 1.49 & 1.39 $\pm$ 0.75 & 2.34 $\pm$ 2.03 & 4.27 $\pm$ 4.41 \\[1ex]
\midrule
\multicolumn{5}{l}{\textbf{Satisfaction Rate (1-5 Likert Scale)}} \\[1ex]
\multicolumn{5}{l}{\textit{All participants}} \\[0.5ex]
Score (Mean $\pm$ SD) & 4.00 $\pm$ 1.08 & 4.16 $\pm$ 0.87 & 3.02 $\pm$ 1.36 & 3.76 $\pm$ 1.08 \\[0.5ex]
\% of Score $\geq$ 4 & 79.6 & 82.0 & 44.9 & 76.0 \\[1ex]
\multicolumn{5}{l}{\textit{Inexperienced participants}} \\[0.5ex]
\# User & 34 & 34 & 33 & 34  \\[0.5ex]
Score  & 4.00 $\pm$ 1.04 & 4.18 $\pm$ 0.87 & 3.15 $\pm$ 1.42 & 4.00 $\pm$ 0.92  \\[1ex]
\multicolumn{5}{l}{\textit{Experienced participants}} \\[0.5ex]
\# User & 15 & 16 & 16 & 16 \\[0.5ex]
Score & 4.00 $\pm$ 1.20 & 4.13 $\pm$ 0.89 & 2.75  $\pm$ 1.24& 3.25 $\pm$ 1.24 \\[1ex]
\midrule
\multicolumn{5}{l}{\textbf{Estimated Time to Manually Create Target Objects from Experts (hours)}} \\[1ex]
\multicolumn{5}{l}{\textit{Expert}} \\[0.5ex]
provided 3D model & 0.4 & 0.4 & 2.1 & -- \\[0.5ex]
not provided 3D model & 2.7 & 4.6 & 6.4 & -- \\[1ex]
\multicolumn{5}{l}{\textit{Novice}} \\[0.5ex]
provided 3D model & 2.0 & 2.0 & 6.0 & -- \\[0.5ex]
not provided 3D model & 7.9 & 10.3 & 14.7 & -- \\[0.5ex]
\bottomrule
\end{tabular}
\label{tab:result_all}
\end{table*}

\begin{figure}
    \centering
    \includegraphics[width=1\linewidth]{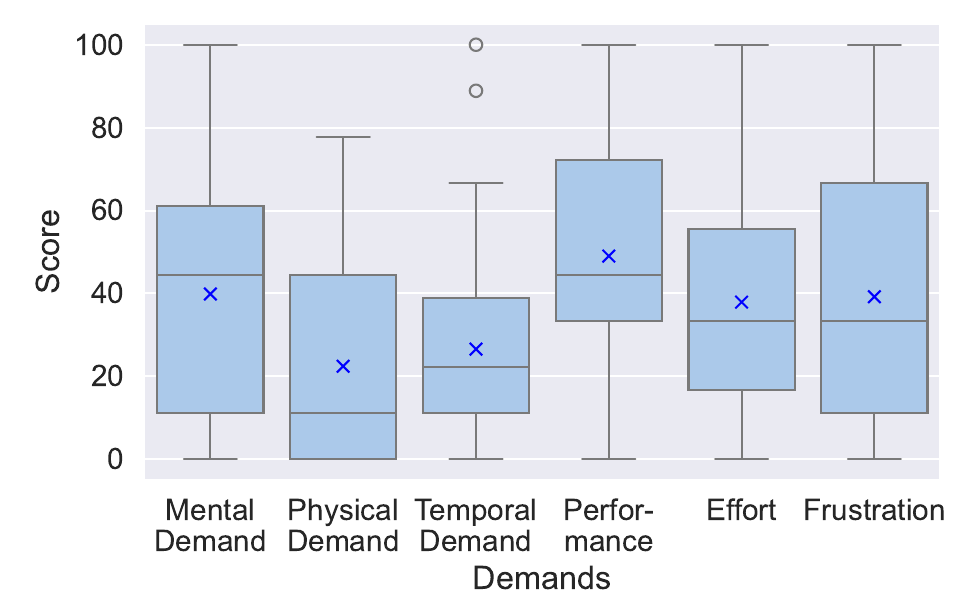}
    \caption{Scores for each of the six fundamental items used to assess task workload in the NASA-TLX.}
    \label{fig:nasa-tlx}
\end{figure}

\paragraph{Task Success Rate} The proportion of participants who completed the upload process after initiating image generation was high for all tasks (Table~\ref{tab:result_all}). Tasks 1 and 2 had a 100~\% success rate, while Tasks 3 and 4 had a 98~\% success rate. These results indicate the usability and reliability of the system in facilitating 3D content creation, even as task complexity increased.

\paragraph{Task Completion Time} We measured the time from each participant's initial image generation to their first successful 3D model upload. Two outliers who took over an hour to complete Tasks 1 and 3, likely due to multitasking or external factors, were excluded from the analysis to maintain data accuracy. Results (Table~\ref{tab:result_all}) show that participants completed each task in approximately 4-7 minutes.

\paragraph{Generation Attempts}
From Fig.~\ref{fig:generation_analysis} and Table~\ref{tab:result_all}, as task complexity increased, the average number of image generation attempts also increased, from 3.32 $\pm$ 2.67 in Task 1 to 5.22 $\pm$ 5.07 in Task 3. This variation indicates that while some users quickly achieved satisfactory results, others engaged in iterative refinement, demonstrating the system's ability to support different user workflows.
In contrast, the number of 3D model generation attempts was 1.84 $\pm$ 1.49 for Task 1 and 2.34 $\pm$ 2.03 for Task 3. This suggests that users are less inclined to iterate on improving the quality of generated 3D models compared to image generation.

\paragraph{User Satisfaction}
Satisfaction was measured using a 5-point Likert scale at the end of each task (Table~\ref{tab:result_all}). The number of respondents varied slightly due to a few non-responses, specifically two in Tasks 1 and 3 and one in Tasks 2 and 4. For participants who provided multiple responses, only the first response was used to ensure data consistency.

Over 75~\% of participants rated their satisfaction as 4 or higher for Tasks 1, 2, and 4. In contrast, Task 3 had a lower mean score of 3.02 $\pm$ 1.36, with only 44.9\% rating their satisfaction as 4 or higher, possibly due to the auto-generated scripts did not work as participants imagined. 

\paragraph{System Usability}
The mean score of SUS was 71.8~$\pm$~15.8, indicating good usability of the developed system.

\paragraph{User Experience Based on Expertise} 
Satisfaction scores showed that inexperienced participants reported higher satisfaction with Task 4 (4.00~$\pm$~0.92) than experienced participants (3.25~$\pm$~1.24), suggesting the system is more supportive for users with less experience. 

To assess the impact of participants' experience in CG production on usability perceptions, we compared SUS scores between users with less than one year of CG experience and those with more than one year, as detailed in Table~\ref{tab:result_all}. In particular, in Task 4, participants with less than one year of experience reported significantly higher satisfaction than their more experienced counterparts (p = 0.034, r = 0.27; Mann-Whitney U test). This indicates that the system is particularly beneficial for novices in CG production.

\paragraph{Task Load Assessment}
Figure \ref{fig:nasa-tlx} shows the score for each item in the NASA-TLX questionnaire collected from the participants. Mental Demand: 39.87~$\pm$~29.91,
Physical Demand: 22.44~$\pm$~24.59,
Temporal Demand: 26.58~$\pm$~22.89,
Performance: 49.02~$\pm$~26.50,
Effort: 37.91~$\pm$~28.39 and
Frustration: 39.22~$\pm$~28.66.
These results provide a nuanced understanding of the task load experienced by users, highlighting that while the system imposes moderate mental and physical demands, it maintains acceptable levels of effort and frustration.
The Raw LTX value, which represents the simple average of demands, was 35.9 across users (SD = 19.0).

\paragraph{System Robustness} 
We measured success rates at all stages of the system: image generation, 3D model generation, and platform upload. Table~\ref{tab:system_robustness} shows these success rates as percentages. The results show that all processing stages maintained success rates above 90~\%, indicating the system's ability to meet diverse user requirements across multiple tasks.

The slightly lower success rate observed in Task 4 can be attributed to its open-ended nature. In this free-form task, users often generated models with vertex counts that exceeded the platform specifications. These complex geometries challenged the automatic mesh reduction algorithms required to meet Cluster's technical constraints, resulting in occasional upload failures.

\begin{table}[htbp]
  \centering
  \caption{Task completion rates (\%)}
  \begin{tabular}{lcccc}
    \textbf{Action} & \textbf{Task 1} & \textbf{Task 2} & \textbf{Task 3} & \textbf{Task 4} \\
    \hline
    Generate Image & 100.00 & 100.00 & 100.00 & 97.89 \\
    Generate GLB & 100.00 & 98.59 & 100.00 & 99.08 \\
    Upload & 91.80 & 98.15 & 98.11 & 71.17 \\
  \end{tabular}
  \label{tab:system_robustness}
\end{table}

\paragraph{API Request Time}
\begin{figure*}
    \centering
    \includegraphics[width=1\linewidth]{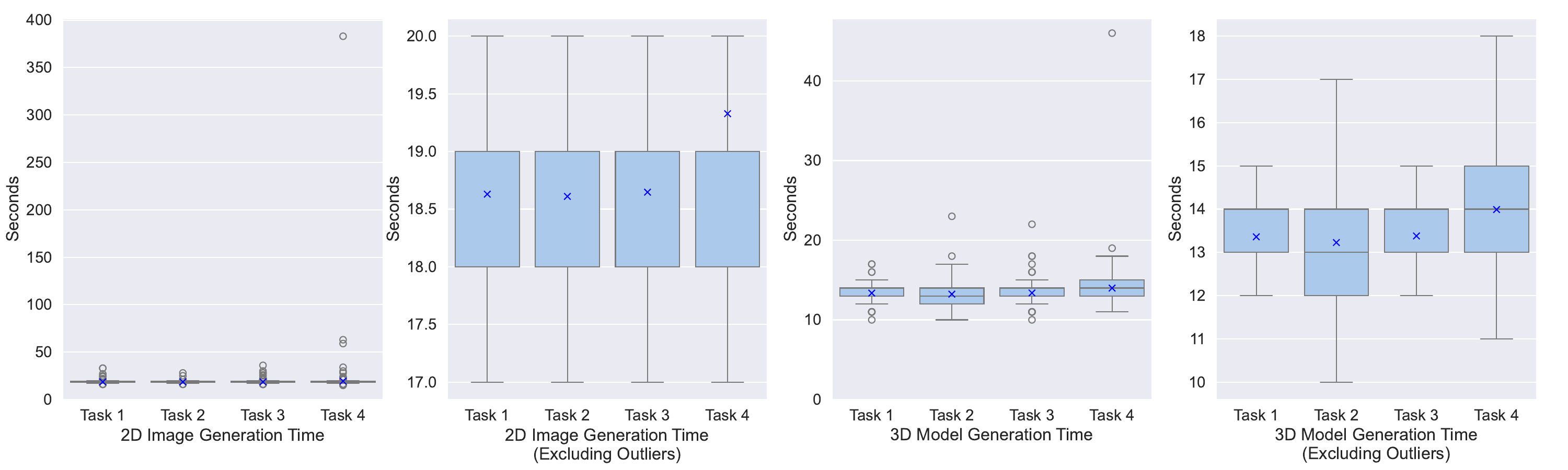}
    \caption{2D image generation time and 3D model generation time for each task. 2D image generation time includes the time taken for LLM calls to expand prompts.}
    \label{fig:generation_time_wo_one_flier}
\end{figure*}

We measured the time taken to generate the images and the time taken to generate the 3D models~(Fig.~\ref{fig:generation_time_wo_one_flier}).
The time taken to generate the image includes the time taken to expand the user's input prompt into a form suitable for image generation using LLM.
In addition, the time taken to call the image generation API is included in the time taken to generate the image.
The time taken to generate the 3D model includes only the time taken to call the 3D model generation API.

Although there are some cases where the generation time is unusually long in the open-ended task 4, the difference in generation time for each task is small, and 2D images and 3D models are generated at a stable speed.

\subsection{Qualitative Feedbacks}
We summarized below the impressions and feedback that participants provided in their free-text descriptions.

\paragraph{Task 1}
Participants appreciated the intuitive process of setting the sitting position and adjusting the size and orientation of the chair, enabling quick creation of practical items. However, they noted that the generated 3D models often suffered from low-resolution textures and distortions, especially when using angled images. Users highlighted the importance of mastering prompt inputs to achieve better results and suggested features like multiple sitting poses and adjustments for incomplete object parts.

\paragraph{Task 2}
Users found the visual adjustment of grip positions intuitive and beneficial, especially for small objects where texture and polygon roughness were less noticeable. However, they reported that mechanical objects such as power drills often resulted in distorted shapes and blurred textures. Suggestions included support for switching between left and right handles, and improvements to texture resolution and model accuracy.

\paragraph{Task 3}
Participants struggled to generate accurate models of complex, symmetrical objects such as small airplanes, often encountering distortions and poor detail. Problems with the generated scripts were common, including unintended motion and abnormal behavior, leading to requests for more granular control over object animation.

\paragraph{Task 4}
Users were impressed with the ability to create a wide variety of items, showing creativity in designing objects such as fountains and seasonal decorations. While they enjoyed the automatic script suggestions, they struggled to reproduce intricate patterns and achieve the intended motion.

\paragraph{Overall Comments}
Participants appreciated the system's ease of use, particularly its ability to quickly transform ideas into tangible forms without requiring prior modeling experience, thereby accelerating world creation and fostering creativity. The automatic scripting and intuitive customization features were well received. However, concerns were raised about the quality of the models and textures generated, especially for detailed or complex objects. Users found that effective scripting required skill, suggesting the need for support mechanisms such as templates or hints. The complexity of the scripting functionality was a barrier for some, indicating the need for more explanation or support systems.

\section{Expert Interviews}
We conducted interviews with professional CG designers to assess the effectiveness of MagicCraft and establish comparative benchmarks for traditional 3D object creation workflows. The expert panel consisted of 7 practitioners (5 male, 2 female; mean age 29~$\pm$~1.8 years) with extensive experience in metaverse content production. The experts represented different professional roles: one motion designer, three CG designers, and three technical artists. They had an average of 3.3~$\pm$~1.3 years of experience in metaverse-specific content creation and 7.6~$\pm$~3.3 years of experience in general CG production.

\subsection{Preliminary Expert Interviews}
The first phase of our expert evaluation focused on setting baseline time requirements for traditional object creation workflows. We presented experts with images of three types of objects generated by MagicCraft: a chair (Task 1), an electric drill (Task 2), and an airplane (Task 3).

Experts estimated the production time requirements for each object type in four different scenarios:
\{expert OR novice\} production time, \{with pre-existing 3D models (GLB/GLTF format) OR modeling from scratch\}.
When scenarios were considered infeasible for an expert or hypothetical novice, the experts identified specific skill requirements that would be necessary to complete the task.

\begin{figure}
    \centering
    \includegraphics[width=1\linewidth]{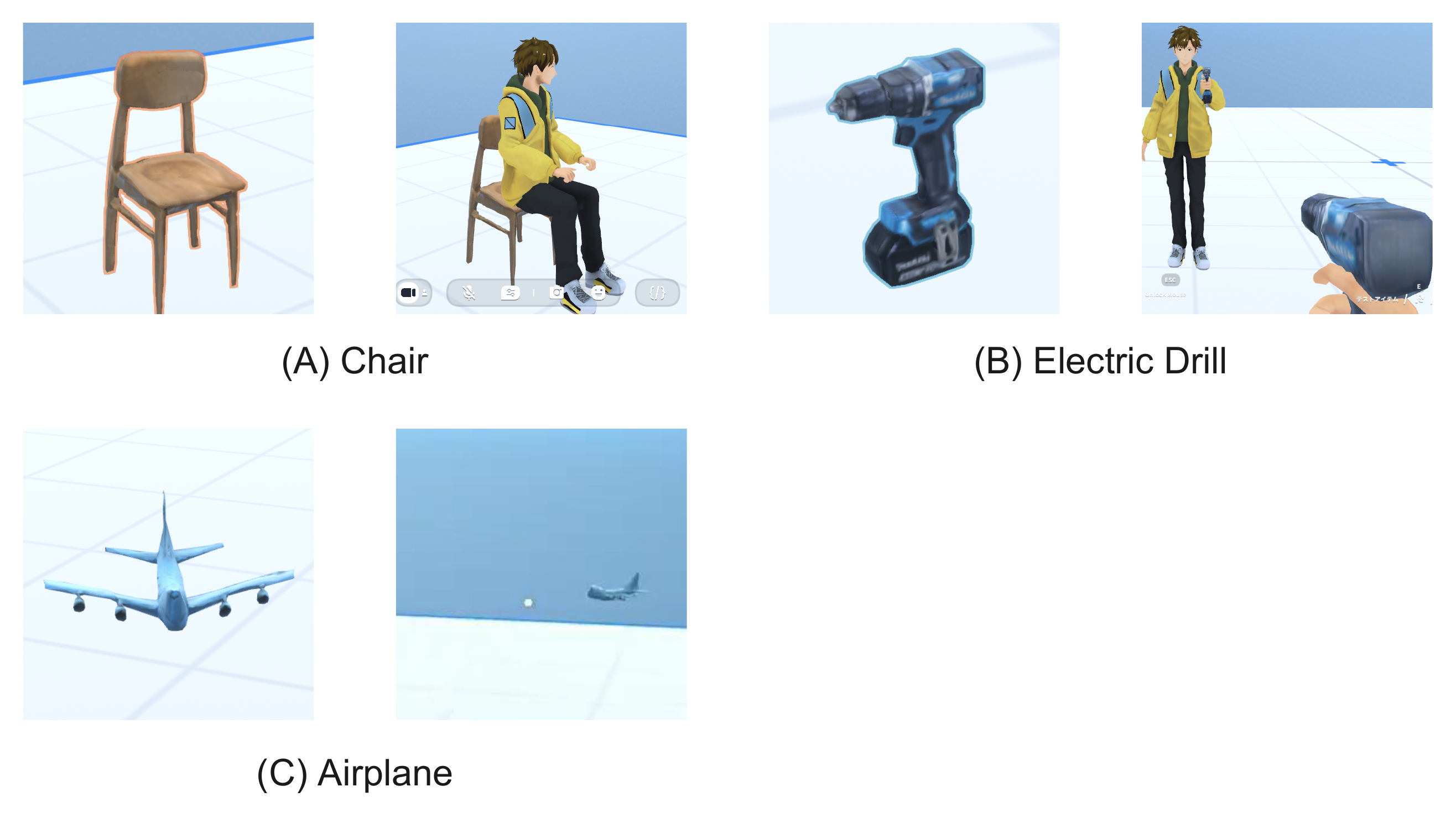}
    \caption{Images presented to CG designers. These are generated by our proposed system for three types of craft items: a chair, an electric drill, and an airplane.}
    \vspace{-5mm}
    \label{fig:preliminary_expert_interview}
\end{figure}

For each scenario, we asked questions about the required production time.
\begin{itemize}
    \item Please estimate the time required to create an object similar to the one shown, assuming a pre-existing 3D model in GLB/GLTF format is available. If certain aspects of this workflow are beyond your expertise, please specify the skills required.
    \item Please estimate the time required to create an object similar to the one shown, including modeling the 3D object from scratch. If any aspects of this workflow are beyond your expertise, please specify the skills required.
\end{itemize}
\subsubsection{Results}
The estimated item creation times reported by the experts are summarized in Table~\ref{tab:result_all}. When existing 3D models were provided, the experts estimated that it would take approximately 0.4 hours to create the chair and the power drill, and 2.1 hours to create the airplane. These times were primarily spent scripting each object to function in the metaverse environment. 

In the absence of existing 3D models, the estimated times increased significantly to 2.7 hours for the chair, 4.6 hours for the power drill, and 6.4 hours for the airplane. Novice CG artists were expected to require much longer times, ranging from 2.0 to 6.0 hours even when using 3D models, and from 7.9 to 14.7 hours when starting from scratch. These results indicate that manual 3D modeling is a time-consuming task, especially for novices and when creating models from scratch.

\subsection{Post-Experiment Qualitative Evaluation of Generated Objects}
\begin{figure*}[t]
    \centering
    \includegraphics[width=1\linewidth]{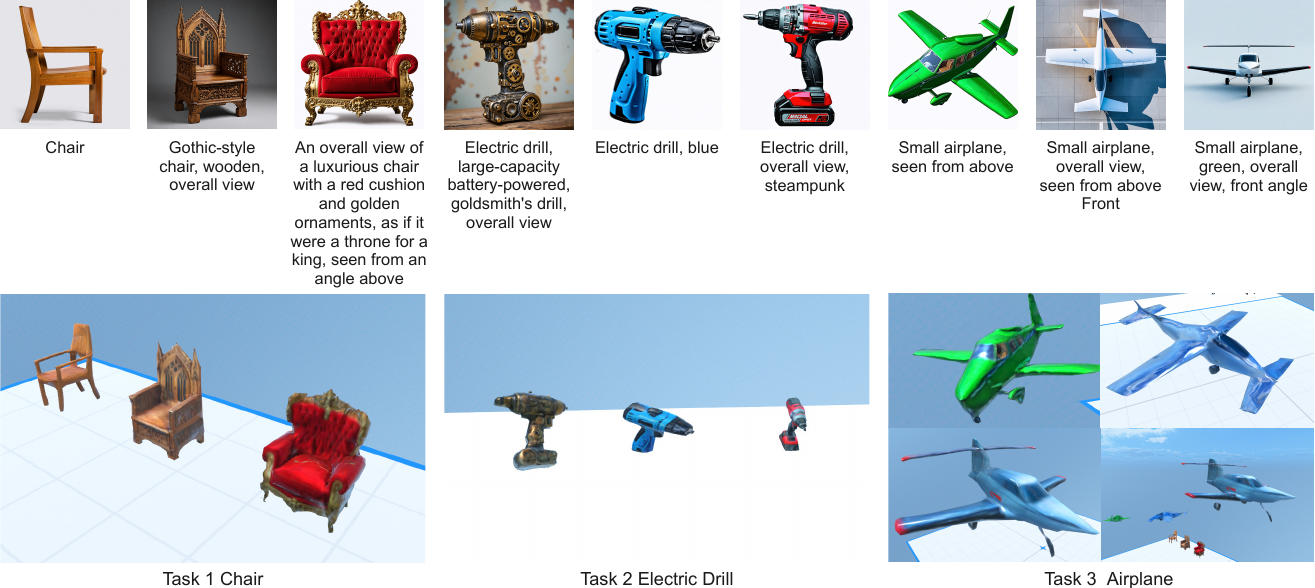}
    \caption{CG designers reviewed these 9 items created by public experiment participants in Task 1, 2 and 3.}
    \label{fig:cg_for_exparts}
\end{figure*}

\begin{figure}[t]
    \centering
    \includegraphics[width=1\linewidth]{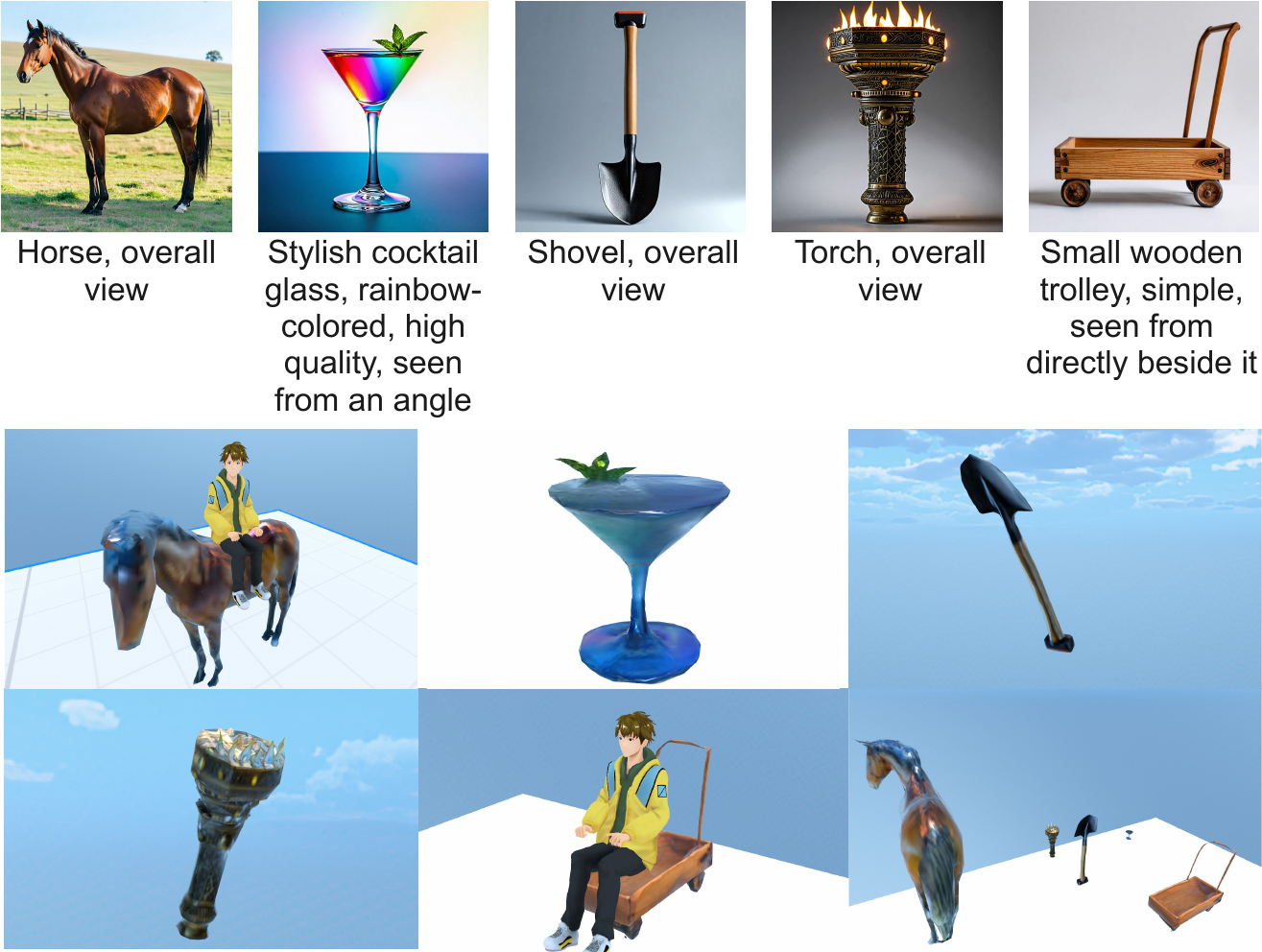}
    \caption{CG designers reviewed these 5 items created by public experiment participants in Task 4 Free Creation.}
    \label{fig:cg_for_exparts_free_creation}
\end{figure}

After the public experiment, we conducted a qualitative evaluation of the objects generated in the public experiment with the help of the experts who participated in the pre-interviews. 

The evaluation was conducted by means of semi-structured interviews. In the interview, 3 objects created by the user in Tasks 1-3 and 5 objects from the freely created examples in Task 4 were selected for evaluation, for a total of 14 objects (Fig.~\ref{fig:cg_for_exparts}).
These objects were selected based on the diversity of viewpoints (front, side, oblique, etc.) and the complexity of textures and shapes in Tasks 1-3, and the diversity of object size, transparency and material gloss in Task 4.

During the semi-structured interviews, a series of questions were asked about each object. The questions were designed to assess various aspects of the objects, including their quality, usability, and creativity.

\subsubsection{Results}
Our expert interviews revealed several critical insights into the performance of MagicCraft in generating functional, dynamic 3D objects from natural language prompts:

\paragraph{Prompt-to-2D Conversion}
Experts found that the generated images generally matched the original textual prompts and successfully depicted objects such as chairs, drills, and airplanes. The images were praised for their resolution, color, and composition. However, some discrepancies were noted: certain objects, such as a "mine cart," appeared more like a wheelbarrow, and objects requiring symmetry, such as airplanes and drills, exhibited distortion or asymmetry. Organic shapes and simpler objects were rendered with greater fidelity, while mechanical and complex items sometimes contained extraneous parts or unnatural distortions.

\paragraph{2D to 3D Conversion}
The transition from 2D images to 3D models maintained moderate accuracy in replicating shapes and textures. Experts observed that organic and less complicated objects, such as horse statues, were effectively modeled. In contrast, mechanical and industrial objects suffered from problems such as lack of straight lines, asymmetry, and low-resolution textures that resulted in poor mesh quality. Models often lacked detail in unseen areas, affecting their plausibility. Scale interpretation was generally appropriate, although some models, such as aircraft, were incorrectly sized for the metaverse environment.

\paragraph{Functionality and Behavior in Metaverse Spaces}
In the metaverse environment, basic interactions such as sitting on chairs or holding objects worked correctly for most models. Experts noted that while basic behaviors were consistent with user intent, there were instances of misalignment and unnatural movement. For example, some airplanes moved diagonally regardless of their orientation, and collider settings did not always match the visual model, causing avatars to intersect objects incorrectly. Script-generated movements were occasionally unnatural or unsmooth, indicating room for improvement in how models respond to the physical laws of the metaverse, such as gravity and collisions.

\paragraph{Overall Impressions and Future Potential}
The experts were impressed with the system's ability to automatically generate usable 3D models from concise natural language prompts. They noted that while the models were not yet ready for formal product deployment due to texture resolution and mesh quality issues, they were of great value for prototyping and environmental testing. The system excelled at creating lifelike and organic objects, but struggled with mechanical or complex designs, likely due to limitations in the AI's understanding of human design intent. Experts also noted that creating effective prompts can be challenging for novices, suggesting a need for improved user guidance. There is enthusiasm for expanding the system's capabilities to include a wider range of objects, such as fantastical designs and creature-like entities, to enhance creativity and expressiveness in metaverse spaces.

\section{Discussion and Future Work}\label{sec:discussion}
\subsection{Efficiency-Quality Trade-offs in Object Creation}
Our evaluation shows that MagicCraft accelerates the creation of 3D objects by a factor of 31 to 164 compared to traditional methods used by professional CG designers. This gain in efficiency is particularly significant given that many of the participants had no prior 3D modeling or programming experience. However, this speed comes at the expense of quality. Expert evaluations revealed that while system-generated objects are effective for prototyping and conceptual exploration, they may not yet meet professional standards for commercial applications due to limitations in texture resolution, mesh quality, and fine detail.

MagicCraft's primary value lies in its ability to facilitate rapid ideation and visualization, addressing a critical bottleneck in the early stages of virtual object development. By lowering technical barriers, the system streamlines creative workflows and potentially reduces production costs. Future development should focus on integrating higher-fidelity generative models and providing advanced editing capabilities to bridge the gap between rapid prototyping and production-quality assets.

Our approach to ensuring successful 3D generation included designing prompts that generated forward-facing object views. The user logs show that the participants typically achieved satisfactory 3D models in 5 or fewer generation attempts (Fig.~\ref{fig:generation_analysis} c), demonstrating the practical usability of the system despite occasional iteration requirements.

\subsection{Avatar-Object Interaction}
User feedback highlighted the effectiveness of our interface to adjust interaction points, particularly for seating configurations. However, grip adjustments for hand-held objects revealed inconsistencies in alignment precision. This challenge stems from the variability of avatar morphologies across metaverse platforms, where user-created avatars often differ significantly from standard models.

Addressing these limitations requires the development of adaptive algorithms that can accommodate different avatar structures and hand dimensions. Potential approaches include machine learning techniques to predict optimal grip points based on avatar characteristics, providing users with more granular control over interaction parameters, and working with platform developers to establish standardized protocols for avatar-object interactions.

\subsection{Spatial Reasoning in Behavioral Scripting}
The generation of dynamic behaviors received lower satisfaction ratings due to unnatural motion patterns and orientation problems. These limitations highlight a fundamental challenge: current large language models lack the robust spatial reasoning capabilities needed to interpret physical properties in 3D environments. LLMs struggle to translate the conceptual understanding of motion into coordinate-specific transformations with appropriate orientation, directionality, and speed.

Improving this capability requires novel approaches to enhance spatial reasoning in generative AI. Promising directions include incorporating spatial context into language model training, developing multimodal systems that process both textual and visual information, and creating intermediate representation formats that bridge natural language and precise 3D transformations. These advances would enable more accurate and complex object behaviors, greatly enhancing the interactivity of virtual environments.

\subsection{Impact for Metaverse Ecosystem}
Patterns of user engagement during our experiments suggest broader implications for metaverse ecosystems. We observed participants creating dedicated virtual spaces to showcase their MagicCraft-generated items, while others explored monetization opportunities within the platform. These behaviors suggest that by lowering the barriers to content creation, MagicCraft empowers users to participate more actively in the metaverse economy.

This democratization of content creation fosters several positive outcomes: increased diversity and volume of user-generated content, enhanced virtual world richness, support for virtual economic systems through digital asset creation and exchange, and increased community engagement through collaborative creation and sharing.

However, these developments raise important considerations regarding intellectual property, content moderation, and platform governance. As AI-generated content enters commercial circulation, questions of ownership attribution, originality verification, and quality assurance become increasingly relevant. Metaverse platforms will need to develop frameworks that address these challenges to ensure fair practices and protect stakeholder interests.

\subsection{Platform Interoperability and Interaction Scope}
Cluster used in our experiments supports only two primary interaction types for specifying object behavior: rideable and grabbable. While these interaction modalities effectively cover most basic operations in current metaverse environments (sitting, mounting vehicles, manipulating tools, etc.), more complex interaction patterns must be implemented through additional scripting. 

Porting this system to alternative metaverse platforms presents significant research opportunities and challenges. Each platform employs different scripting architectures, asset processing pipelines, and interaction frameworks, requiring adaptive implementation strategies to maintain functional consistency across ecosystems.

\section{Conclusion}
We introduced MagicCraft, a system that allows users to generate functional and dynamic 3D objects for metaverse platforms from natural language prompts. By integrating LLM with image generation, image-to-3D conversion, and automated behavior scripting, MagicCraft enables users with no prior experience in 3D modeling or programming to create complex, interactive objects in virtual environments.
We implemented MagicCraft on the Cluster platform and conducted a thorough evaluation, including a comparative study with seven expert CG designers and an online experiment with 51 general users. The results showed that MagicCraft significantly reduced the time required to create 3D objects and lowered the technical skill barriers. Users successfully created a variety of objects, customized their attributes, and deployed them in metaverse spaces.

This paper has highlighted both the capabilities and limitations of applying current LLM and generative AI to metaverse platforms for dynamic 3D object generation. We hope that our work will inspire future research to further improve the integration of generative AI into the metaverse.

\bibliographystyle{IEEEtran}
\bibliography{main}

\begin{thebibliography}{10}
\providecommand{\url}[1]{#1}
\csname url@samestyle\endcsname
\providecommand{\newblock}{\relax}
\providecommand{\bibinfo}[2]{#2}
\providecommand{\BIBentrySTDinterwordspacing}{\spaceskip=0pt\relax}
\providecommand{\BIBentryALTinterwordstretchfactor}{4}
\providecommand{\BIBentryALTinterwordspacing}{\spaceskip=\fontdimen2\font plus
\BIBentryALTinterwordstretchfactor\fontdimen3\font minus \fontdimen4\font\relax}
\providecommand{\BIBforeignlanguage}[2]{{%
\expandafter\ifx\csname l@#1\endcsname\relax
\typeout{** WARNING: IEEEtran.bst: No hyphenation pattern has been}%
\typeout{** loaded for the language `#1'. Using the pattern for}%
\typeout{** the default language instead.}%
\else
\language=\csname l@#1\endcsname
\fi
#2}}
\providecommand{\BIBdecl}{\relax}
\BIBdecl

\bibitem{vrchat}
{VRChat Inc.}, ``\BIBforeignlanguage{en}{{VRChat}},'' \url{https://hello.vrchat.com/}, 2024, accessed: 2024-6-15.

\bibitem{roblox}
{Roblox Corporation}, ``Roblox,'' \url{https://www.roblox.com/}, 2024, accessed: 2024-6-15.

\bibitem{clsuter}
{Cluster, Inc.}, ``Cluster,'' \url{https://cluster.mu/}, 2024, accessed: 2024-6-15.

\bibitem{rombach2022stable}
\BIBentryALTinterwordspacing
R.~Rombach, A.~Blattmann, D.~Lorenz, P.~Esser, and B.~Ommer, ``High-resolution image synthesis with latent diffusion models,'' in \emph{2022 IEEE/CVF Conference on Computer Vision and Pattern Recognition (CVPR)}.\hskip 1em plus 0.5em minus 0.4em\relax Los Alamitos, CA, USA: IEEE Computer Society, jun 2022, pp. 10\,674--10\,685. [Online]. Available: \url{https://doi.ieeecomputersociety.org/10.1109/CVPR52688.2022.01042}
\BIBentrySTDinterwordspacing

\bibitem{poole2023dreamfusion}
\BIBentryALTinterwordspacing
B.~Poole, A.~Jain, J.~T. Barron, and B.~Mildenhall, ``Dreamfusion: Text-to-3d using 2d diffusion,'' in \emph{The Eleventh International Conference on Learning Representations}, 2023. [Online]. Available: \url{https://openreview.net/forum?id=FjNys5c7VyY}
\BIBentrySTDinterwordspacing

\bibitem{OpenAI_GPT4_2023}
\BIBentryALTinterwordspacing
OpenAI, ``Gpt-4 technical report,'' \emph{ArXiv}, vol. abs/2303.08774, 2023. [Online]. Available: \url{https://arxiv.org/abs/2303.08774}
\BIBentrySTDinterwordspacing

\bibitem{Giunchi2024-gt}
D.~Giunchi, N.~Numan, E.~Gatti, and A.~Steed, ``{DreamCodeVR}: Towards democratizing behavior design in virtual reality with {Speech-Driven} programming,'' in \emph{2024 {IEEE} Conference Virtual Reality and {3D} User Interfaces ({VR})}.\hskip 1em plus 0.5em minus 0.4em\relax IEEE, Mar. 2024, pp. 579--589.

\bibitem{kurai-2025-magicitem}
R.~Kurai, T.~Hiraki, Y.~Hiroi, Y.~Hirao, M.~Perusquía-Hernández, H.~Uchiyama, and K.~Kiyokawa, ``Magicitem: Dynamic behavior design of virtual objects with large language models in a commercial metaverse platform,'' \emph{IEEE Access}, pp. 1--1, 2025.

\bibitem{kurai-2025-IEEEVR-poster}
------, ``An implementation of magiccraft: Generating interactive 3d objects and their behaviors from text for commercial metaverse platforms,'' in \emph{2025 {IEEE} on Virtual Reality and {3D} User Interfaces Abstracts and Workshops ({VRW})}, 2025, to appear.

\bibitem{neos}
{Solirax}, ``Neos,'' \url{https://neos.com/}, 2024, accessed: 2024-6-15.

\bibitem{resonite}
{Yellow Dog Man Studios}, ``Resonite,'' \url{https://resonite.com/}, 2024, accessed: 2024-6-15.

\bibitem{Ondrejka2004}
C.~Ondrejka, ``Escaping the gilded cage: User created content and building the metaverse,'' \emph{New York Law School Law Journal}, vol.~49, 05 2004.

\bibitem{blender}
{Blender Foundation}, ``Blender,'' \url{https://www.blender.org/}, 2024, accessed: 2024-6-15.

\bibitem{unity}
{Unity Technologies}, ``\BIBforeignlanguage{ja}{Unity},'' \url{https://unity.com/}, 2024, accessed: 2024-6-15.

\bibitem{unreal-engine}
{Epic Games, Inc.}, ``\BIBforeignlanguage{en}{Unreal engine},'' \url{https://www.unrealengine.com/}, 2024, accessed: 2024-6-15.

\bibitem{cho2015physical}
\BIBentryALTinterwordspacing
Y.~H. Cho, S.~Y. Yim, and S.~Paik, ``Physical and social presence in 3d virtual role-play for pre-service teachers,'' \emph{The Internet and Higher Education}, vol.~25, pp. 70--77, 2015. [Online]. Available: \url{https://www.sciencedirect.com/science/article/pii/S1096751615000032}
\BIBentrySTDinterwordspacing

\bibitem{Latoschik2017}
\BIBentryALTinterwordspacing
M.~E. Latoschik, D.~Roth, D.~Gall, J.~Achenbach, T.~Waltemate, and M.~Botsch, ``The effect of avatar realism in immersive social virtual realities,'' in \emph{Proceedings of the 23rd ACM Symposium on Virtual Reality Software and Technology}, ser. VRST '17, 2017. [Online]. Available: \url{https://doi.org/10.1145/3139131.3139156}
\BIBentrySTDinterwordspacing

\bibitem{Pan2017}
\BIBentryALTinterwordspacing
Y.~Pan and A.~Steed, ``The impact of self-avatars on trust and collaboration in shared virtual environments,'' \emph{PLOS ONE}, vol.~12, no.~12, pp. 1--20, Dec. 2017. [Online]. Available: \url{https://doi.org/10.1371/journal.pone.0189078}
\BIBentrySTDinterwordspacing

\bibitem{Hindmarsh2000}
\BIBentryALTinterwordspacing
J.~Hindmarsh, M.~Fraser, C.~Heath, S.~Benford, and C.~Greenhalgh, ``Object-focused interaction in collaborative virtual environments,'' \emph{ACM Transactions on Computer-Human Interaction}, vol.~7, no.~4, p. 477–509, dec 2000. [Online]. Available: \url{https://doi.org/10.1145/365058.365088}
\BIBentrySTDinterwordspacing

\bibitem{bowman2008ui}
D.~Bowman, S.~Coquillart, B.~Froehlich, M.~Hirose, Y.~Kitamura, K.~Kiyokawa, and W.~Stuerzlinger, ``3d user interfaces: New directions and perspectives,'' \emph{IEEE computer graphics and applications}, vol.~28, pp. 20--36, 11 2008.

\bibitem{tang2023dreamgaussian}
J.~Tang, J.~Ren, H.~Zhou, Z.~Liu, and G.~Zeng, ``Dreamgaussian: Generative gaussian splatting for efficient 3d content creation,'' \emph{arXiv preprint arXiv:2309.16653}, 2023.

\bibitem{bensadoun2024meta3dgen}
\BIBentryALTinterwordspacing
R.~Bensadoun, T.~Monnier, Y.~Kleiman, F.~Kokkinos, Y.~Siddiqui, M.~Kariya, O.~Harosh, R.~Shapovalov, B.~Graham, E.~Garreau, A.~Karnewar, A.~Cao, I.~Azuri, I.~Makarov, E.-T. Le, A.~Toisoul, D.~Novotny, O.~Gafni, N.~Neverova, and A.~Vedaldi, ``Meta 3d gen,'' 2024. [Online]. Available: \url{https://arxiv.org/abs/2407.02599}
\BIBentrySTDinterwordspacing

\bibitem{SF3D-Boss2024-uz}
M.~Boss, Z.~Huang, A.~Vasishta, and V.~Jampani, ``{SF3D}: Stable fast {3D} mesh reconstruction with {UV}-unwrapping and illumination disentanglement,'' \emph{arXiv [cs.CV]}, Aug. 2024.

\bibitem{lin2023magic3d}
C.-H. Lin, J.~Gao, L.~Tang, T.~Takikawa, X.~Zeng, X.~Huang, K.~Kreis, S.~Fidler, M.-Y. Liu, and T.-Y. Lin, ``Magic3d: High-resolution text-to-3d content creation,'' in \emph{IEEE Conference on Computer Vision and Pattern Recognition ({CVPR})}, 2023.

\bibitem{Xie2022}
Y.~Xie, T.~Takikawa, S.~Saito, O.~Litany, S.~Yan, N.~Khan, F.~Tombari, J.~Tompkin, V.~Sitzmann, and S.~Sridhar, ``Neural fields in visual computing and beyond,'' \emph{Computer Graphics Forum}, 2022.

\bibitem{huynhthe2023ai}
\BIBentryALTinterwordspacing
T.~Huynh-The, Q.-V. Pham, X.-Q. Pham, T.~T. Nguyen, Z.~Han, and D.-S. Kim, ``Artificial intelligence for the metaverse: A survey,'' \emph{Engineering Applications of Artificial Intelligence}, vol. 117, p. 105581, 2023. [Online]. Available: \url{https://www.sciencedirect.com/science/article/pii/S0952197622005711}
\BIBentrySTDinterwordspacing

\bibitem{zhang2022motiondiffuse}
M.~Zhang, Z.~Cai, L.~Pan, F.~Hong, X.~Guo, L.~Yang, and Z.~Liu, ``Motiondiffuse: Text-driven human motion generation with diffusion model,'' \emph{arXiv preprint arXiv:2208.15001}, 2022.

\bibitem{De_La_Torre2024-be}
F.~De~La~Torre, C.~M. Fang, H.~Huang, A.~Banburski-Fahey, J.~Amores~Fernandez, and J.~Lanier, ``{LLMR}: Real-time prompting of interactive worlds using large language models,'' in \emph{Proceedings of the {CHI} Conference on Human Factors in Computing Systems}, ser. CHI '24, May 2024.

\bibitem{Pearce2022}
H.~Pearce, B.~Ahmad, B.~Tan, B.~Dolan-Gavitt, and R.~Karri, ``Asleep at the keyboard? assessing the security of github copilot's code contributions,'' in \emph{Proceedings of the 43rd IEEE Symposium on Security and Privacy}, 2022, pp. 754--768.

\bibitem{koyama2022bo}
\BIBentryALTinterwordspacing
Y.~Koyama and M.~Goto, ``Bo as assistant: Using bayesian optimization for asynchronously generating design suggestions,'' in \emph{Proceedings of the 35th Annual ACM Symposium on User Interface Software and Technology}, ser. UIST '22.\hskip 1em plus 0.5em minus 0.4em\relax New York, NY, USA: Association for Computing Machinery, 2022. [Online]. Available: \url{https://doi.org/10.1145/3526113.3545664}
\BIBentrySTDinterwordspacing

\bibitem{Chen2021}
M.~Chen, J.~Tworek, H.~Jun, Q.~Yuan, H.~P. de~Oliveira~Pinto, J.~Kaplan, H.~Edwards, Y.~Burda, N.~Joseph, G.~Brockman, A.~Ray, R.~Puri, G.~Krueger, M.~Petrov, H.~Khlaaf, G.~Sastry, P.~Mishkin, B.~Chan, S.~Gray, N.~Ryder, M.~Pavlov, A.~Power, L.~Kaiser, M.~Bavarian, C.~Winter, P.~Tillet, F.~P. Such, D.~Cummings, M.~Plappert, F.~Chantzis, E.~Barnes, A.~Herbert-Voss, W.~H. Guss, A.~Nichol, A.~Paino, N.~Tezak, J.~Tang, I.~Babuschkin, S.~Balaji, S.~Jain, W.~Saunders, C.~Hesse, A.~N. Carr, J.~Leike, J.~Achiam, V.~Misra, E.~Morikawa, A.~Radford, M.~Knight, M.~Brundage, M.~Murati, K.~Mayer, P.~Welinder, B.~McGrew, D.~Amodei, S.~McCandlish, I.~Sutskever, and W.~Zaremba, ``Evaluating large language models trained on code,'' \emph{ArXiv}, 2021.

\bibitem{github-copilot}
{GitHub, Inc.}, ``{GitHub} copilot,'' \url{https://copilot.github.com/}, 2024, accessed: 2024-6-15.

\bibitem{Vaithilingam2022}
\BIBentryALTinterwordspacing
P.~Vaithilingam, T.~Zhang, and E.~L. Glassman, ``Expectation vs experience: Evaluating the usability of code generation tools powered by large language models,'' in \emph{Extended Abstracts of the 2022 CHI Conference on Human Factors in Computing Systems}, ser. CHI EA '22.\hskip 1em plus 0.5em minus 0.4em\relax New York, NY, USA: Association for Computing Machinery, 2022. [Online]. Available: \url{https://doi.org/10.1145/3491101.3519665}
\BIBentrySTDinterwordspacing

\bibitem{ahmad-etal-2021-unified}
\BIBentryALTinterwordspacing
W.~Ahmad, S.~Chakraborty, B.~Ray, and K.-W. Chang, ``Unified pre-training for program understanding and generation,'' in \emph{Proceedings of the 2021 Conference of the North American Chapter of the Association for Computational Linguistics: Human Language Technologies}, K.~Toutanova, A.~Rumshisky, L.~Zettlemoyer, D.~Hakkani-Tur, I.~Beltagy, S.~Bethard, R.~Cotterell, T.~Chakraborty, and Y.~Zhou, Eds.\hskip 1em plus 0.5em minus 0.4em\relax Online: Association for Computational Linguistics, Jun. 2021, pp. 2655--2668. [Online]. Available: \url{https://aclanthology.org/2021.naacl-main.211}
\BIBentrySTDinterwordspacing

\bibitem{GLTF-The-Khronos(r)-3D-Formats-Working-GroupUnknown-uc}
{The Khronos® 3D Formats Working Group}, ``{glTF™} 2.0 specification,'' \url{https://registry.khronos.org/glTF/specs/2.0/glTF-2.0.html}, accessed: 2024-9-18.

\bibitem{GPT-4o-mini}
OpenAI, ``\BIBforeignlanguage{en}{{GPT}-{4o} mini: advancing cost-efficient intelligence},'' \url{https://openai.com/index/gpt-4o-mini-advancing-cost-efficient-intelligence/}, accessed: 2024-9-19.

\bibitem{GRADIO-Abid2019-sr}
A.~Abid, A.~Abdalla, A.~Abid, D.~Khan, A.~Alfozan, and J.~Zou, ``Gradio: Hassle-free sharing and testing of {ML} models in the wild,'' \emph{arXiv [cs.LG]}, Jun. 2019.

\bibitem{Docker-Merkel2014-iw}
D.~Merkel, ``Docker: lightweight linux containers for consistent development and deployment,'' \emph{Linux J.}, 2014.

\bibitem{Cloud-Run}
{Google Cloud}, ``\BIBforeignlanguage{en}{Cloud run},'' \url{https://cloud.google.com/run/?hl=en}, accessed: 2024-9-19.

\bibitem{UGPT-4o}
OpenAI, ``\BIBforeignlanguage{en}{Hello {GPT}-{4o}},'' \url{https://openai.com/index/hello-gpt-4o/}, accessed: 2024-9-19.

\bibitem{Stability-AI-Models}
{Stability AI}, ``\BIBforeignlanguage{en}{Stability {AI} image models —},'' \url{https://stability.ai/stable-image}, accessed: 2024-9-19.

\bibitem{Brooke1996-kw}
\BIBentryALTinterwordspacing
J.~Brooke, ``{SUS}: A 'quick and dirty' usability scale,'' in \emph{Usability Evaluation In Industry}.\hskip 1em plus 0.5em minus 0.4em\relax CRC Press, Jun. 1996, pp. 207--212. [Online]. Available: \url{http://dx.doi.org/10.1201/9781498710411-35}
\BIBentrySTDinterwordspacing

\bibitem{Nasa1986-vw}
\BIBentryALTinterwordspacing
{NASA}, ``Nasa task load index ({TLX}) v.1.0 manual,'' 1986. [Online]. Available: \url{http://humansystems.arc.nasa.gov/groups/TLX/downloads/TLX.pdf}
\BIBentrySTDinterwordspacing

\end{thebibliography}

\newpage

\begin{IEEEbiography}[{\includegraphics[width=1in,height=1.25in,clip,keepaspectratio]{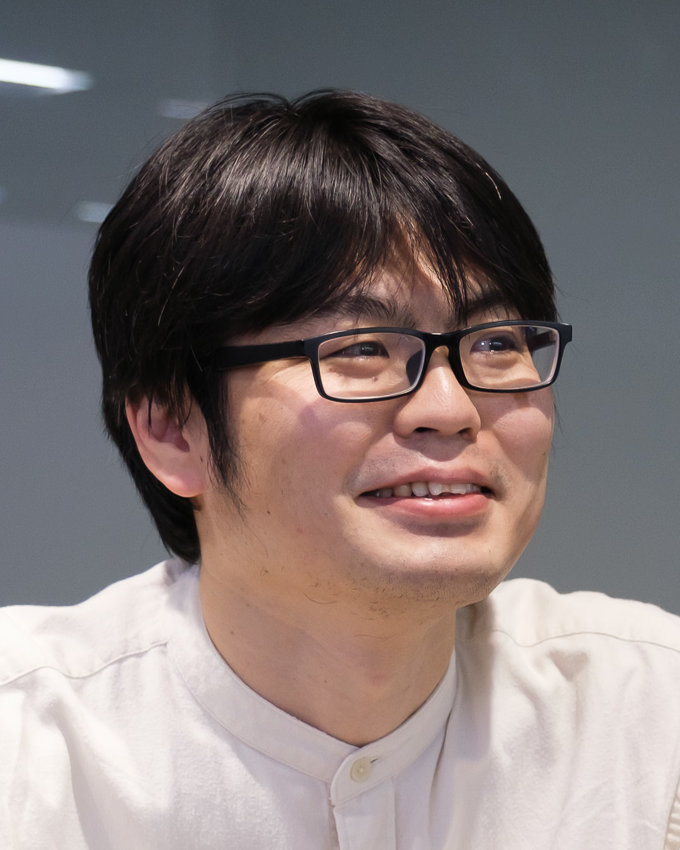}}]{Ryutaro Kurai} (Student Member, IEEE) received his B.S. and M.S. degrees from Hokkaido University, Japan in 2007 and 2009, respectively. He spent 2 years as a researcher at ERATO Minato Discrete Structure Manipulation System Project (2013-2015). Currently, he is an engineering manager at Cluster, Inc. His research interests include metaverse, human-computer interaction, large language model, and software engineering.
\end{IEEEbiography}

\begin{IEEEbiography}[{\includegraphics[width=1in,height=1.25in,clip,keepaspectratio]{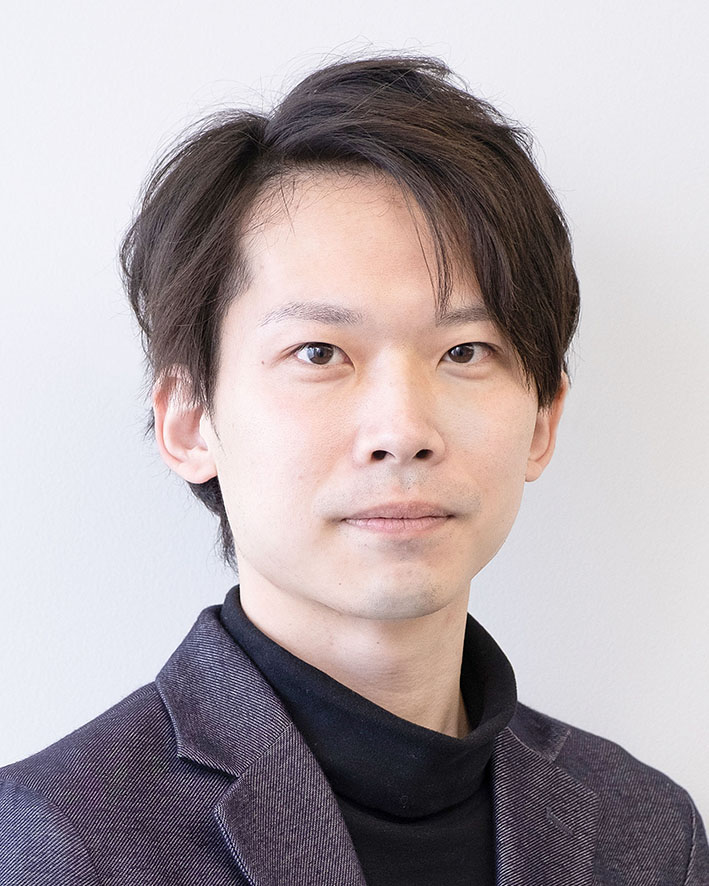}}]{Takefumi Hiraki} (Member, IEEE) received his B.S., M.S., and Ph.D. degrees from the University of Tokyo, Japan in 2014, 2016, and 2019, respectively. From 2019, he worked as a Japan Society for the Promotion of Science (JSPS) Research Fellow at the Graduate School of Engineering Science at Osaka University. Since 2021, he served as an assistant professor at the Institute of Library, Information and Media Science at University of Tsukuba. Currently, he is a senior research scientist at Cluster Metaverse Lab at Cluster, Inc. and a collaborative associate professor at the University of Electro-Communications. His research interests include augmented reality, metaverse, human-computer interaction, haptic interfaces, and soft robotics.
\end{IEEEbiography}


\begin{IEEEbiography}[{\includegraphics[width=1in,height=1.25in,clip,keepaspectratio]{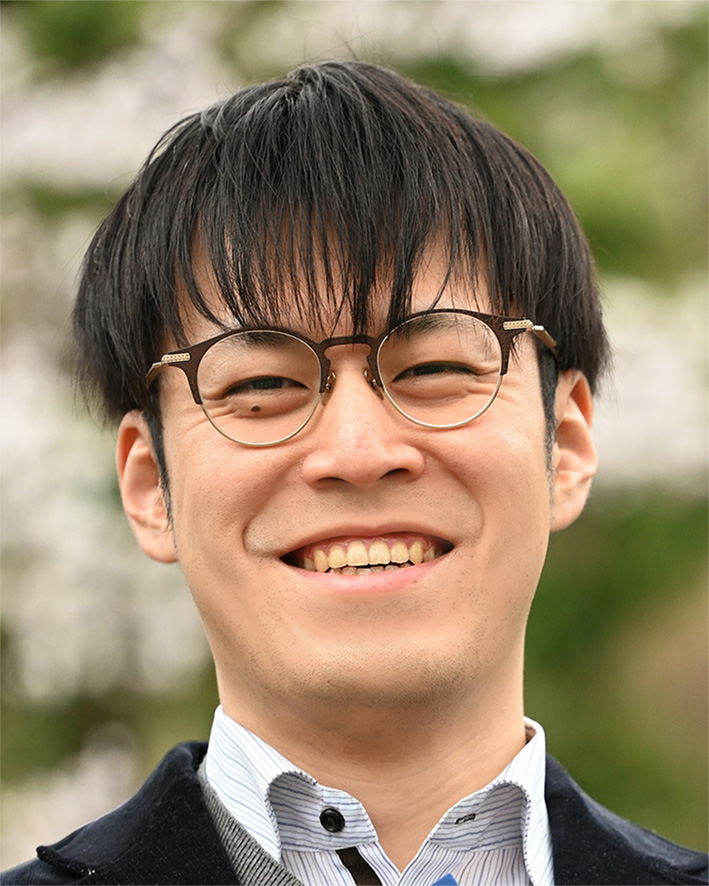}}]{Yuichi Hiroi} (Member, IEEE) received his BS and MS degrees in computer sciences from Keio University, Japan in 2015 and 2017, and received his Ph.D. degrees from Tokyo Institute of Technology, Japan in 2022. He also spent two years as a researcher at NTT Service Evolution Laboratories (2017-2018). From 2022, he worked as a Japan Society for the Promotion of Science (JSPS) Research Fellow at the University of Tokyo. 
Currently, he is a senior research scientist in Cluster Metaverse Lab, Japan, from 2023. His research interests mainly include virtual and augmented reality, head-mounted displays, and vision augmentation.
\end{IEEEbiography}

\begin{IEEEbiography}[{\includegraphics[width=1in,height=1.25in,clip,keepaspectratio]{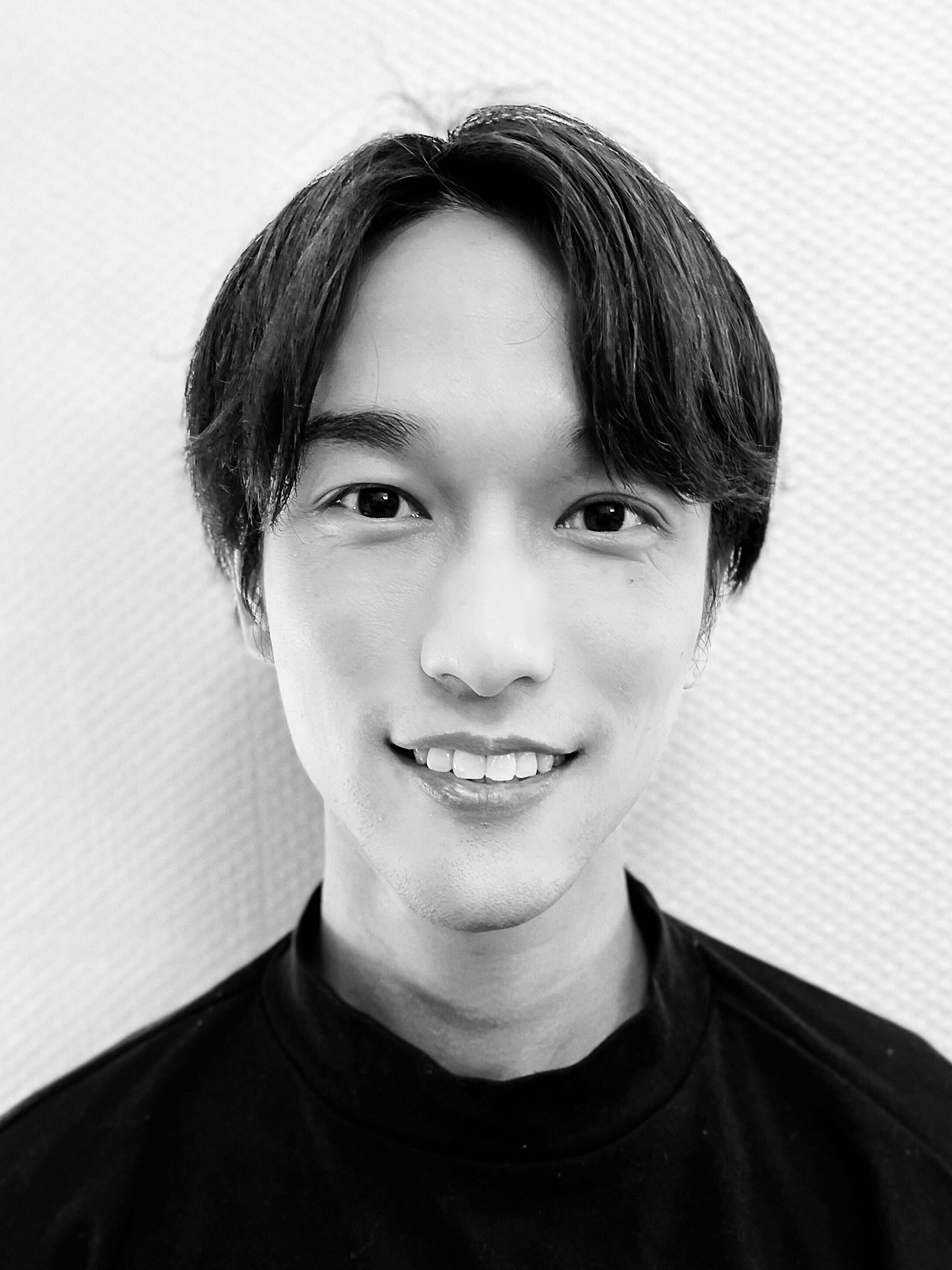}}]{Yutaro Hirao} received his B.S. and M.S. in engineering from Waseda University (14-18, 18-20) in Japan, and his Ph.D. in information science and technology from the University of Tokyo (20-23). He is an Assistant Professor at Nara Institute of Science and Technology (NAIST), Japan (23-). His main research interests include virtual reality (VR), cross-modal interaction, embodiment, and haptics.
\end{IEEEbiography}

\begin{IEEEbiography}[{\includegraphics[width=1in,height=1.25in,clip,keepaspectratio]{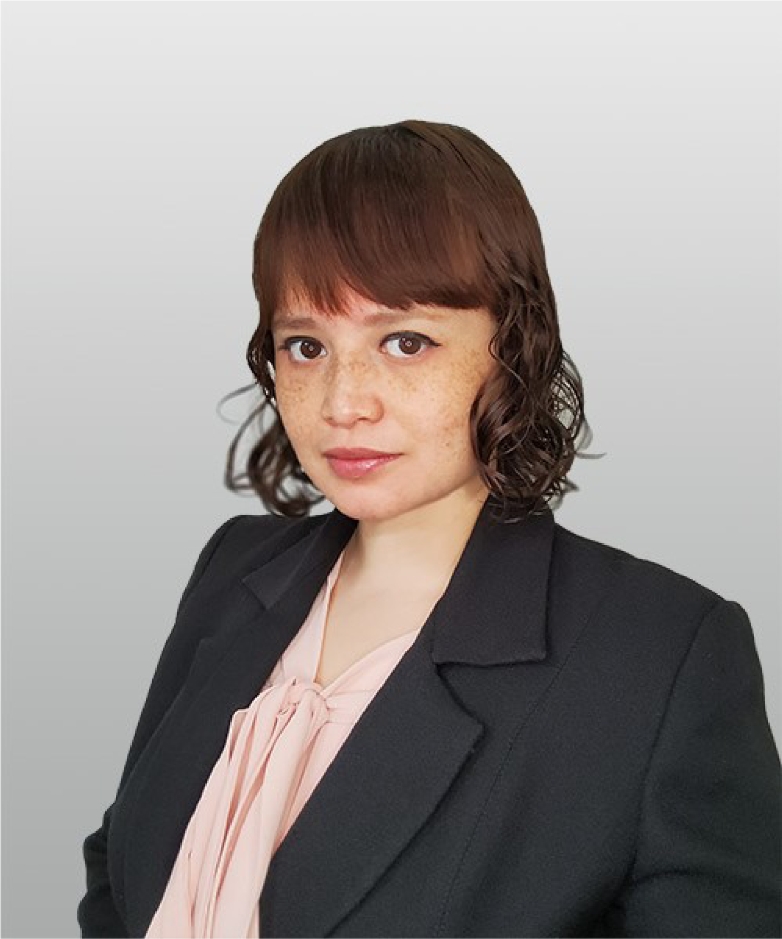}}]{Monica Perusqu\'{i}a-Hern\'{a}ndez} received her BSc in electronic systems engineering (2009) from the Instituto Tecnol\'{o}gico y de Estudios Superiores de Monterrey, Mexico; her MSc in human-technology interaction (2012) and the professional doctorate in engineering in user-system interaction (2014) from the Eindhoven University of Technology, the Netherlands. In 2018, she obtained her Ph.D. in Human Informatics from the University of Tsukuba, Japan. She is an Assistant Professor at the Nara Institute of Science and Technology. Her research interests include affective computing, biosignal processing, augmented human technology, and artificial intelligence.
\end{IEEEbiography}

\begin{IEEEbiography}[{\includegraphics[width=1in,height=1.25in,clip,keepaspectratio]{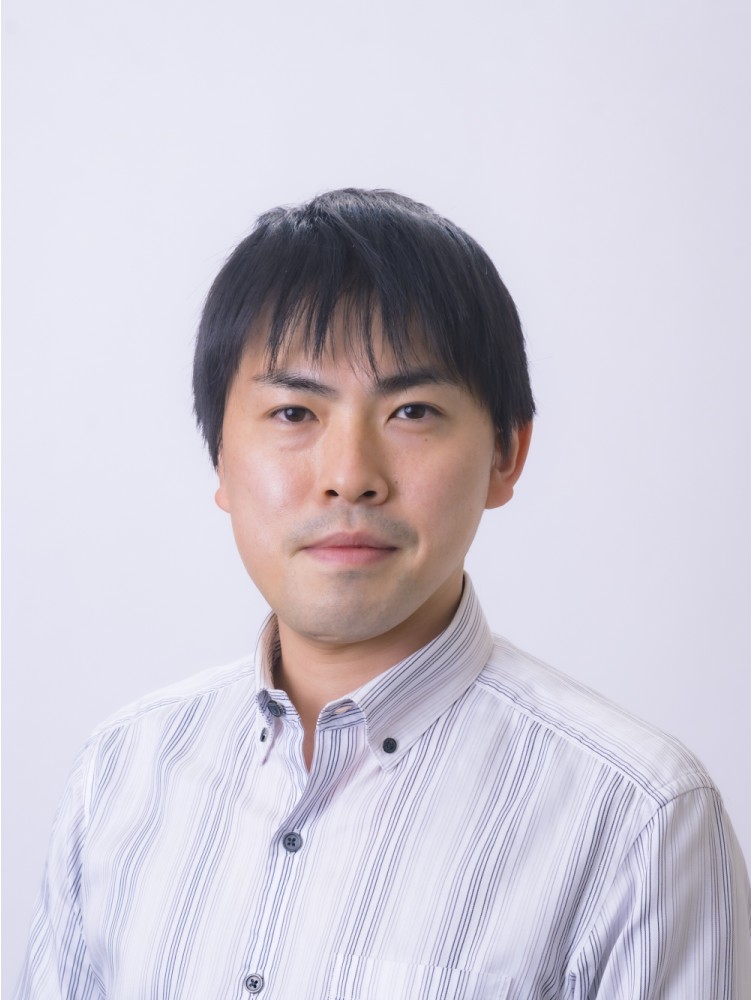}}]{Hideki Uchiyama} received his Ph.D. in engineering from Keio University in 2010. After working at INRIA, Toshiba, and Kyushu University, he is an Associate Professor at Nara Institute of Science and Technology. His research interests include computer vision, inertial sensing, and their applications.
\end{IEEEbiography}

\begin{IEEEbiography}[{\includegraphics[width=1in,height=1.25in,clip,keepaspectratio]{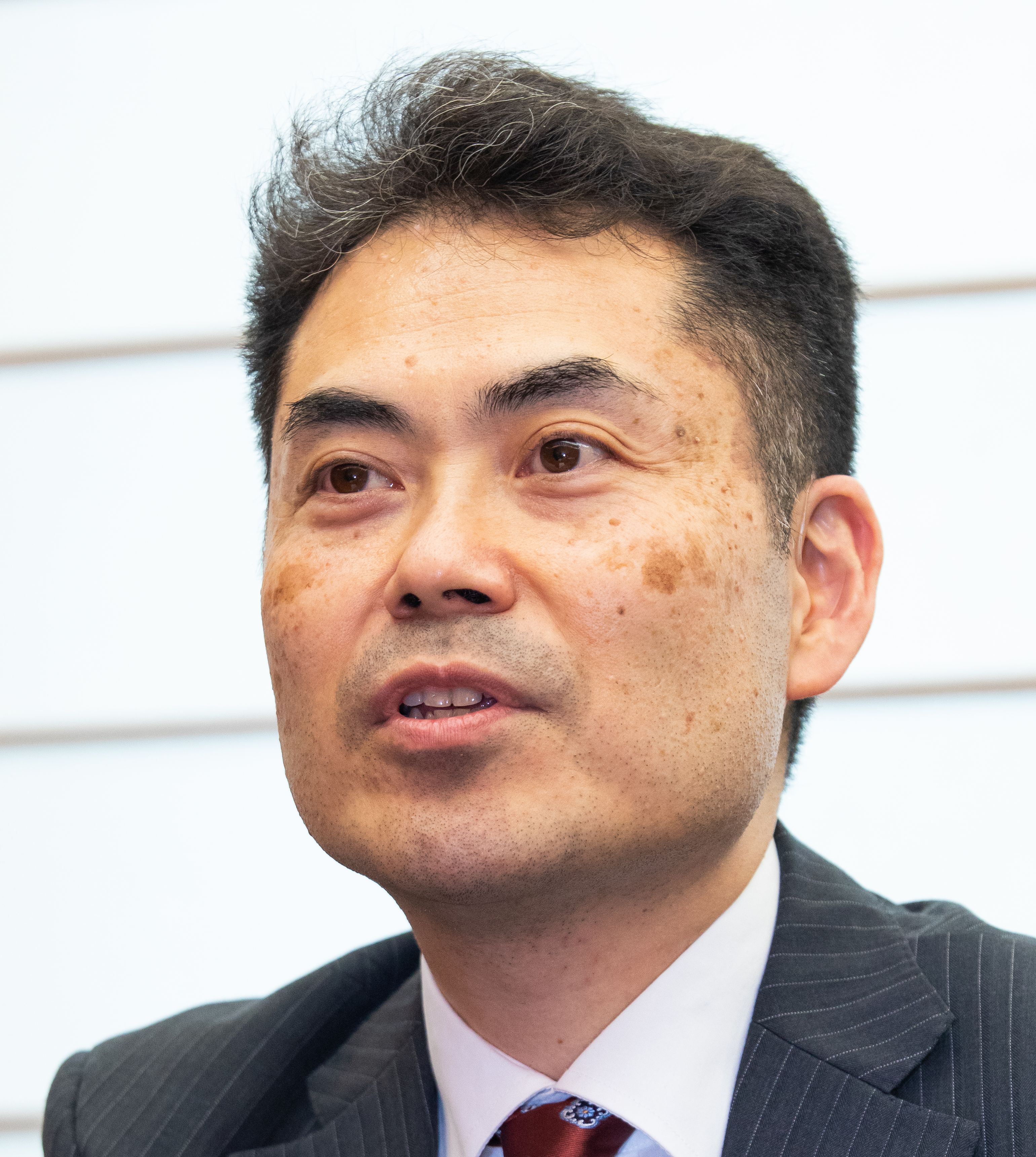}}]{Kiyoshi Kiyokawa} (Member, IEEE) received his Ph.D. degree in information systems from Nara Institute of Science and Technology (NAIST) in 1998. After working at the National Institute of Information and Communications Technology (NICT) and Osaka University, he is currently a Professor at NAIST, leading the Cybernetics and Reality Engineering Laboratory. His research interests include virtual reality, augmented reality, human augmentation, 3D user interfaces, CSCW, and context awareness. He is also a Fellow of the Virtual Reality Society of Japan. He is an associate editor-in-chief of IEEE TVCG and an inductee of the IEEE VGTC Virtual Reality Academy (Inaugural Class).
\end{IEEEbiography}

\EOD

\end{document}